\documentclass[sigconf,natbib=false]{acmart}
\PassOptionsToPackage{protrusion, expansion}{microtype}
\PassOptionsToPackage{style=ieee, citestyle=numeric, sortcites=true, backend=bibtex}{biblatex}
\PassOptionsToPackage{scaled=.9}{inconsolata}
    
\usepackage{biblatex}

\usepackage{comment}
\usepackage{amsmath, amsthm, mathtools, bm} 
    
\usepackage{graphicx, xcolor}               
    \definecolor{BLUE}{rgb}{.0, .2, .6}     
    \definecolor{BLUE2}{HTML}{1e50a2}       
    \definecolor{RED}{HTML}{c9171e}         
    \definecolor{RED2}{HTML}{D7003A}        
	\definecolor{INK}{HTML}{595857}         
	\definecolor{YELLOW}{HTML}{f1c40f}      
	\definecolor{GREEN}{rgb}{0,0.6,0}       
	\definecolor{MIDGRAY}{rgb}{0.5,0.5,0.5} 

    \definecolor{B}    {HTML}{2b66d3}
    \definecolor{B2}   {HTML}{003399}
    \definecolor{R}    {HTML}{c9171e}
    \definecolor{R2}   {HTML}{d7003a}
    \definecolor{INK}  {HTML}{595857}
    \definecolor{Y}    {HTML}{f1c40f}
    \definecolor{G}    {HTML}{009a00}
    \definecolor{GRAY} {HTML}{808080}
    \definecolor{MAUVE}{HTML}{9400D1}
\usepackage{url, hyperref}                  
	\hypersetup{
	   pdfborder={0 0 0}}
\usepackage{adjustbox}                      
\usepackage{booktabs, multirow}             
\usepackage{caption, subcaption}            
\usepackage{enumitem}                       
\usepackage{lipsum}                         
\usepackage{setspace}                       
\usepackage[boxed]{algorithm}       				
\floatstyle{plaintop}
\usepackage{algpseudocode}
    \algrenewcommand{\alglinenumber}[1]{{\scriptsize\bfseries\ttfamily\color{RED}#1}}

\makeatletter
\newcommand\fs@ruled@nomiddle{\def\@fs@cfont{\bfseries}\let\@fs@capt\floatc@ruled
    \def\@fs@pre{}%
    \def\@fs@mid{\vskip 1pt\hrule height.8pt depth0pt \kern2pt}%
    \def\@fs@post{\vskip 5pt\hrule height.8pt depth0pt \relax}%
  \let\@fs@iftopcapt\iftrue}
\renewcommand\fst@algorithm{\fs@ruled@nomiddle}
\makeatother
    
\usepackage{xpatch}
    \makeatletter
    \xpatchcmd{\algorithmic}{\ALG@tlm\z@}{\ALG@tlm\z@\leftmargin 10pt}{}{}
    \makeatother
\usepackage{footnote}                       
\usepackage{listings}						
\usepackage{colortbl}
\usepackage{tikz}
	\newcommand*\Circled[1]{
		\tikz[baseline=(char.base)]{%
			\node[shape=circle, draw=none, fill=gray!40, thick, inner sep=0.6pt] (char) {%
				\textcolor{black}{\sffamily#1}}; }}
				
\PassOptionsToPackage{protrusion, expansion}{microtype} 

\usepackage{soul}

\usepackage{array}
\usepackage{rotating}

\newcommand{\SEC}{\textcolor{black}{\S}}
\newcommand{\FIG}{\textcolor{black}{Figure}}
\newcommand{\TAB}{\textcolor{black}{Table}}

\newcommand{\behold}[1]{\hl{\strut #1}}

\def\dataTableBase#1#2#3{%
    \sffamily\begin{tabular}{@{}#3@{}}\scshape\color{gray}#1\\[-.8ex]#2\end{tabular}%
}

\def\datasetName#1#2{%
    \dataTableBase{#1}{#2}{l}%
}

\def\datumSize#1#2{%
    \dataTableBase{\scriptsize#1}{#2}{r}%
}

\def\datasetField#1#2{%
    \dataTableBase{\scriptsize\normalfont\sffamily#1 in total}{\ttfamily #2}{r}%
}

\def\tableTitleFormat{%
    \bfseries\scshape\sffamily%
}

\newcommand\grandTableTitle[3]{%
    \sffamily\begin{tabular}{@{}#1@{}}\scshape\sffamily\bfseries#2\\[-.8ex]\scshape#3\end{tabular}%
}

\newcommand{\TabUnitStyle}{\scshape\bfseries\sffamily\color{gray}}

\newcommand\singleLineTableHeadFormat{%
    \bfseries\scshape\sffamily%
}

\newcommand\doubleLineTableHead[3]{%
    \sffamily\scshape\bfseries\begin{tabular}{@{}#1@{}}#2\\[-.8ex]#3\end{tabular}%
}

\newcommand{%
    \tableHeadBold}[1]{\multicolumn{1}{c}{\sffamily\bfseries\begin{tabular}{@{}c@{}}#1\end{tabular}}%
}

\newcommand{%
    \notAvailable}{{\setlength{\fboxsep}{1pt}\fbox{$\times$}}%
}

\copyrightyear{2020}
\acmYear{2020}
\setcopyright{rightsretained}
\acmConference[PACT '20]{Proceedings of the 2020 International
Conference on Parallel Architectures and Compilation
Techniques}{October 3--7, 2020}{Virtual Event, GA, USA}
\acmBooktitle{Proceedings of the 2020 International Conference on
Parallel Architectures and Compilation Techniques (PACT '20), October
3--7, 2020, Virtual Event, GA, USA}
\acmDOI{10.1145/3410463.3414624}
\acmISBN{978-1-4503-8075-1/20/10}

\hypersetup{pdfauthor={J. Tian}, pdftitle={cuSZ}}

\renewcommand{\behold}[1]{#1}

\newcommand\cuszToCpuMin{\behold{242.9$\times$}}
\newcommand\cuszToCpuMax{\behold{370.1$\times$}}
\newcommand\cuszToOmpMin{\behold{11.0$\times$}}
\newcommand\cuszToOmpMax{\behold{13.1$\times$}}
\newcommand\cuszToCuzfpBrMin{\behold{2.41$\times$}}
\newcommand\cuszToCuzfpBrMax{\behold{3.48$\times$}}

\newcommand{\termPQ}{{predict-quant}}
\newcommand{\termDQ}{{\scshape dual-quant}}
\newcommand{\termPREQ}{{\scshape prequant}}
\newcommand{\termPOSTQ}{{\scshape postquant}}

\title[\textsc{cuSZ}]{\textsc{cuSZ}: An Efficient GPU-Based Error-Bounded Lossy Compression Framework for Scientific Data}
\newcommand{\wsu}{Washington State University}
\newcommand{\ua}{The University of Alabama}
\newcommand{\ucr}{University of California, Riverside}
\newcommand{\clemson}{Clemson University}
\newcommand{\anl}{Argonne National Laboratory}
\newcommand{\ornl}{Oak Ridge National Laboratory}

\newcommand{\AFFIL}[3]{%
    \affiliation{%
        \institution{\small #1}
    }
    }

\newcommand{\WSU}{\AFFIL{\wsu}{Pullman}{WA}}
\newcommand{\UA}{\AFFIL{\ua}{Tuscaloosa}{AL}}
\newcommand{\UCR}{\AFFIL{\ucr}{Riverside}{CA}}
\newcommand{\ANL}{\AFFIL{\anl}{Lemont}{IL}}
\newcommand{\ORNL}{\AFFIL{\ornl}{Oak Ridge}{TN}}
\newcommand{\CLEMSON}{\AFFIL{\clemson}{Clemson}{SC}}

\author{Jiannan Tian}{\WSU}
\email{jiannan.tian@wsu.edu}

\author{Sheng Di}{\ANL}
\email{sdi1@anl.gov}

\author{Kai Zhao}{\UCR}
\email{kzhao016@ucr.edu}

\author{Cody Rivera}{\UA}
\email{cjrivera1@crimson.ua.edu}

\author{Megan Hickman Fulp}{\CLEMSON}
\email{mlhickm@g.clemson.edu}

\author{Robert Underwood}{\CLEMSON}
\email{robertu@g.clemson.edu}

\author{Sian Jin}{\WSU}
\email{sian.jin@wsu.edu}

\author{Xin Liang}{\ORNL}
\email{liangx@ornl.gov}

\author{Jon Calhoun}{\CLEMSON}
\email{jonccal@clemson.edu}

\author{Dingwen Tao}
\authornote{Corresponding author: Dingwen Tao (\url{dingwen.tao@wsu.edu}), School of EECS, Washington State University, Pullman, WA 99164, USA.}
{\WSU}
\email{dingwen.tao@wsu.edu}

\author{Franck Cappello}{\ANL}
\email{cappello@mcs.anl.gov}

\renewcommand{\shortauthors}{Tian et~al.}

\begin{abstract}
    Error-bounded lossy compression is a state-of-the-art data reduction 
    technique for HPC applications because it not only significantly reduces storage overhead but also can retain high fidelity for postanalysis.
    Because supercomputers and HPC applications are becoming heterogeneous using accelerator-based architectures, in particular GPUs, several development teams have recently released GPU versions of their lossy compressors.
    However, existing state-of-the-art GPU-based lossy compressors suffer from either low compression and decompression throughput or low compression quality.
    In this paper, we present an optimized GPU version, \textsc{cuSZ}, for one of the best error-bounded lossy compressors---SZ. To the best of our knowledge, \textsc{cuSZ} is the \textsc{first error-bounded} lossy compressor on GPUs for scientific data. 
    Our contributions are fourfold.
    (1) We propose a \textsc{dual-quantization} scheme to entirely remove the data dependency in the prediction step of SZ such that this step can be performed very efficiently on GPUs.
    (2) We develop an efficient customized Huffman coding for the SZ compressor on GPUs.
    (3) We implement \textsc{cuSZ} using CUDA and optimize its performance by improving the utilization of GPU memory bandwidth.
    (4) We evaluate our \textsc{cuSZ} on five real-world HPC application datasets from the Scientific Data Reduction Benchmarks and compare it with other state-of-the-art methods on both CPUs and GPUs. Experiments show that our \textsc{cuSZ} improves SZ's compression throughput by up to {\cuszToCpuMax} and \cuszToOmpMax, respectively, over the production version running on single and multiple CPU cores, respectively, while getting the same quality of reconstructed data. It also improves the compression ratio by up to {\cuszToCuzfpBrMax} on the tested data compared with another state-of-the-art GPU supported lossy compressor. 
\end{abstract}



\begin{document}

\maketitle


\section{Introduction}
\label{sec:introduction}

Large-scale high-performance computing (HPC) applications can generate extremely large volumes of scientific data.
For instance, the Hardware/Hybrid Accelerated Cosmology Code (HACC)~\cite{hacc} can simulate 1$\sim$10 trillion particles in one simulation and produce up to 220 TB of data per snapshot, bringing up a total of 22 PB of data during the simulation~\cite{miraio} with only one hundred timesteps/snapshots. Such a large volume of data is imposing an unprecedented burden on supercomputer storage and interconnects~\cite{liang2018error} for both storing data to persistent storage and loading data for postanalysis and visualization. Therefore, data reduction has attracted great attention from HPC application users for reducing the volumes of data to be moved to/from storage systems. The common approaches are simply decimating snapshots periodically and adopting an interpolation for data reconstruction. However, such approaches result in a significant loss of valuable information for postanalysis~\cite{liang-decimation-drbsd4}. Traditional data deduplication and lossless compression have also been used for shrinking data size but suffer from very limited reduction ratios on HPC floating-point datasets. Specifically, deduplication generally reduces the size of scientific datasets by only 20\% to 30\%~\cite{meister2012study}, and lossless compression achieves a reduction ratio of up to about 2:1 in general~\cite{son2014data}. This is far from scientists' desired compression ratios, which are around 10:1 or higher (such as Community Earth Simulation Model (CESM)~\cite{baker2014methodology}).

Error-bounded lossy compression has been proposed to significantly reduce data size while ensuring acceptable data distortion for users~\cite{sz17}. SZ~\cite{sz16, sz17} is a state-of-the-art error-bounded lossy compression framework for scientific data (to be detailed in \SEC\ref{sec:sz-background}), which often offers higher compression qualities (or better rate distortions) than other state-of-the-art techniques~\cite{liang2018error}. However, as illustrated in prior work~\cite{sz16, sz17}, SZ suffers from low compression and decompression throughput, which is only tens to hundreds of megabytes per second on a single CPU core. This throughput is far from enough for extreme-scale applications or advanced instruments with extremely high data acquisition rates, which is a major concern for corresponding users. 
The LCLS-II laser~\cite{lcls}, for instance, may produce  data at a rate of 250 GB/s~\cite{use-case-Franck}, such that corresponding researchers require an extremely fast compression solution that can still have relatively high compression ratios---for example, 10:1---with preserved data accuracy.
In order to match such a high data production rate, leveraging multiple graphics processing units (GPUs) is a fairly attractive solution because of its massive single-instruction multiple-thread (SIMT) mechanism and its high programmability as opposed to FPGAs or ASICs~\cite{tian2020wavesz}. 
Moreover, the SZ algorithm follows $\mathcal{O}(n)$ time complexity and employs large amounts of read and write operations in the memory, and hence its performance is eventually bounded by memory bandwidth.
State-of-the-art GPUs cannot only provide high computation capability but also provide high memory bandwidth.
For example, NVIDIA V100 GPU can provide at least one higher order magnitude of memory bandwidth than state-of-the-art CPUs can~\cite{nv100}.

SZ, however, cannot be run on GPUs efficiently because of the lack of parallelism in its design.
The main challenges are twofold: \Circled{1} the tight dependency in the prediction-quan\-tiza\-tion step of the SZ algorithm incurs expensive synchronizations across iterations in a GPU implementation; and \Circled{2} during the customized Huffman coding step of the SZ algorithm, coding and decoding each symbol based on the constructed Huffman tree involve many different branches (see \SEC\ref{sec:sz-background} for more details). This process causes serious warp divergence and random memory access issues,
which inevitably lead to low GPU memory bandwidth utilization and performance.

To solve these issues, this paper presents an optimized GPU version of the SZ algorithm, called \textsc{cuSZ}, and proposes a series of optimization techniques for \textsc{cuSZ} to achieve high compression and decompression throughputs on GPUs.
Specifically, we focus on the main performance bottlenecks (Lorenzo prediction~\cite{ibarria2003out} and customized Huffman coding~\cite{sz17}) and improve their performance for GPUs. 
We propose a novel technique called \textsc{dual-quantization} that can be applied to any prediction-based compression algorithms to alleviate the tight dependency in its prediction step.
Moreover, according to prior work~\cite{use-case-Franck}, a strict error-controlling scheme of lossy compression is needed by many HPC applications for their scientific explorations and postanalyses. However, the state-of-the-art GPU-based lossy compressors such as cuZFP~\cite{cuZFP} are not error-bounded. 
To the best of our knowledge, \textsc{cuSZ}~\footnote{The code is available at \url{https://github.com/hipdac-lab/cuSZ}.} is \textsc{the first strictly error-bounded lossy compressor on GPU for scientific data.} 
Our contributions are summarized as follows.
\begin{itemize}[noitemsep, topsep=2pt, leftmargin=1.3em]
    \item We propose a generic \textsc{dual-quantization} scheme to entirely remove the data dependencies in the prediction-quantization step of lossy compression and apply it to Lorenzo predictor in SZ algorithm.
    \item We develop an efficient customized Huffman coding for SZ on GPUs with fine- and coarse-grained parallelism.
    \item We carefully implement \textsc{cuSZ} and optimize its performance on CUDA architecture. In particular, we fine-tune the chunk size in Huffman coding and develop an adaptive method that selects 32-bit or 64-bit representation dynamically for Huffman code and can significantly improve GPU memory bandwidth utilization. 
    \item We evaluate our proposed \textsc{cuSZ} on five real-world HPC application datasets provided by a public repository, \emph{Scientific Data Reduction Benchmarks}~\cite{sdrbench}, and compare it with other state-of-the-art methods on both CPUs and GPUs. Experiments show that the \textsc{cuSZ} can significantly improve both compression throughput by up to {\cuszToCpuMax} and {\cuszToOmpMax} over the production version of SZ running on single CPU core and multiple CPU cores, respectively. \textsc{cuSZ} has up to {\cuszToCuzfpBrMax} higher compression ratio than another advanced GPU supported lossy compressor  with reasonable data distortion.
\end{itemize}

The rest of the paper is organized as follows. In \SEC\ref{sec:sz-background}, we discuss the SZ lossy compression in detail. In \SEC\ref{sec:design}, we propose our novel optimizations for the GPU version of SZ and implement it using CUDA. In \SEC\ref{sec:evaluation}, we present the evaluation results based on five real-world simulation datasets from the Scientific Data Reduction Benchmarks and compare \textsc{cuSZ} with other state-of-the-art compressors on both CPU and GPU. In \SEC\ref{sec:related}, we discuss related work. In \SEC\ref{sec:conclusion}, we present our conclusions and discuss our future work.

\section{SZ Background}
\label{sec:sz-background}

\begin{figure*}[ht]
    \includegraphics[width=\linewidth]{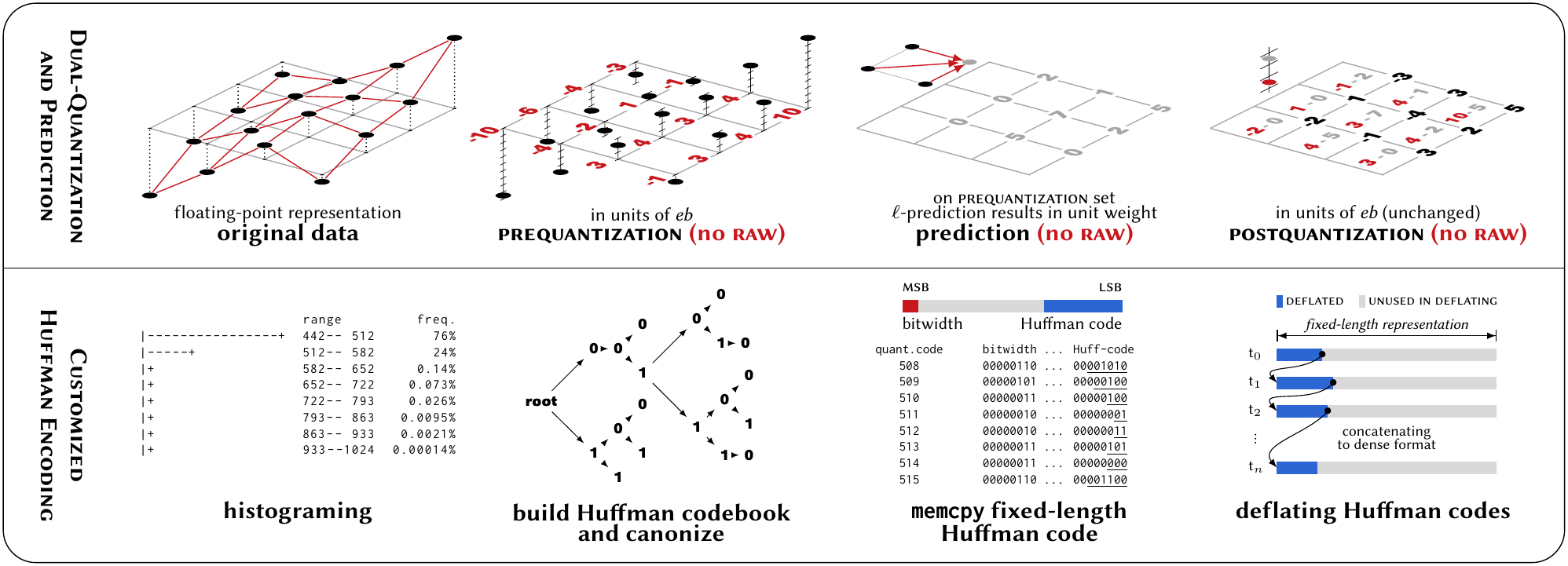}
    \caption{The system overview of \textsc{cuSZ}. The top 4 figures illustrate a {\termDQ} example, which has no loop-carried RAW. The bottom 4 figures correspond to the four subprocedures of our customized Huffman coding described in \SEC\ref{sub:huffman}. }
    \label{fig:overview}
    \vspace{-4mm}
\end{figure*}

Many scientific applications require a strict error-bounded control when using lossy compression to achieve accurate postanalysis and visualization for scientific discovery, as well as a high compression ratio. SZ~\cite{sz16,sz17} is a prediction-based lossy compression framework designed for scientific data that strictly controls the global upper bound of compression error. Given a user-set error bound $eb$, SZ guarantees $|\,d\!-\!d^\bullet|\!<\!eb$, where $d$ and $d^\bullet$ are the original value and the decompressed value, respectively. SZ's algorithm involves five key steps: preprocessing, data prediction, linear-scaling quantization, customized variable-length encoding, and optional lossless compression, e.g., \texttt{gzip}~\cite{gzip} and \texttt{Zstd}~\cite{zstd}. 
\begin{enumerate}
    [noitemsep, topsep=2pt, leftmargin=1.3em, label=\textbf{\arabic*})]
    \item \textbf{Preprocessing} SZ performs a preprocessing step, such as linearization in version 1.0 or a logarithmic transform for the pointwise relative error bound in version 2.0~\cite{xincluster18}.
    \item \textbf{Data Prediction} SZ predicts the value of each data point by a data-fitting predictor, e.g., a \textit{Lorenzo} predictor~\cite{ibarria2003out} (abbreviated as $\ell$-predictor) based on its neighboring values. In order to guarantee that the compression error is always within the user-set error bound, the predicted values must be exactly the same in between the compression procedure and decompression procedure.
    To this end, the neighbor values used in the prediction have to be the decompressed values instead of the original values. 
    \item \textbf{Linear-Scaling Quantization} SZ computes the difference between the predicted value and original value for each data point and performs a linear-scaling quantization~\cite{sz17} to convert the difference to an integer based on the user-set error bound.
    \item \textbf{Customized Variable-Length Coding} SZ adopts a customized Huffman coding to reduce the data size significantly, because the integer codes generated by the linear-scaling quantization are likely distributed unevenly, especially when the data are mostly predicted accurately.
    \item \textbf{Lossless Compression} The last step optionally further compresses the encoded data by a lossless compressor such as \texttt{Zstd}~\cite{zstd}, which may significantly reduce the size due to potential repeated patterns in the bitstream.
\end{enumerate}

In this work, we focus mainly on the SZ compressor, because much prior work~\cite{foster2017computing,di2018efficient,liang2018error,pastri,understand-compression-ipdps18,jin2019deepsz,liang2019significantly,zhao2020significantly} has verified that SZ yields the best compression quality among all  the prediction-based compressors. However, it is nontrivial to port SZ on GPUs because of the strict constraints in its compression design. For instance, the data used in the prediction must be updated by decompressed values, such that the data prediction in the SZ compressor~\cite{sz16,sz17} needs to be performed one by one in a sequential order. 
This requirement introduces a loop-carried \emph{read-after-write} (RAW) dependency during the compression (will be discussed in \SEC\ref{par:raw}), making SZ hard to parallelize.

We mainly focus on  SZ-1.4 instead of SZ-2.0 because the 2.0 model is particularly designed for  low-precision use cases with visualization goals, in which the compression ratio can reach up to several hundred while the reconstructed data often have large data distortions. Recent studies~\cite{use-case-Franck}, however, demonstrate that scientists often require a relatively high precision (or low error bound) for their sophisticated postanalysis beyond visualization purposes. In this situation (with relatively low error bounds), SZ-2.0 has very similar (or even slightly worse, if not for all the cases) compression qualities to those of  SZ-1.4, as demonstrated in~\cite{liang2018error}. Accordingly, our design for the GPU-accelerated SZ lossy compression is based on SZ-1.4 and takes advantage of both algorithmic and GPU hardware characteristics. Moreover, the current CPU version of SZ does not support SIMD vectorization and has no specific improvement on the arithmetic performance. Therefore, the CPU baseline used in our following evaluation is based on the nonvectorized single-core and multicore implementation.

\section{Design Methodology of \textsc{cuSZ}}
\label{sec:design}

In this section, we propose our novel lossy compression design, \textsc{cuSZ}, for CUDA architectures based on the SZ model.
A system overview of our proposed \textsc{cuSZ} is shown in \FIG~\ref{fig:overview}.
We develop different coarse- and fine-grained parallelization techniques to each subprocedure in compression and decompression.
Specifically, we first employ a data-chunking technique to exploit coarse-grained data parallelism. The chunking technique is used throughout the whole \textsc{cuSZ} design, including lossless (step 2 and 3) and lossy (step 1, 4, and 5) procedures in both compression and decompression. We then deeply analyze the RAW data dependency in SZ and propose a novel {two-phase prediction-quantization approach}, namely, \textsc{dual-quantization}, which totally eliminates the data dependency in the prediction-quantization (abbreviated as {\termPQ}) step. Furthermore, we provide an in-depth breakdown analysis of Huffman coding and develop an efficient Huffman coding on GPUs with multiple optimizations. 
A summary of our parallelization techniques is shown in \TAB~\ref{tab:how-to-par}. 

\begin{table}[ht]
    \vskip-1ex
    \centering\footnotesize\sffamily
    \newcommand{\YES}{$\bullet$}
\newcommand{\TabTitle}[1]{\multicolumn{1}{@{}c@{}}{\rotatebox{90}{#1}}}
\renewcommand{\arraystretch}{1.2}
\begin{tabular}{@{}r|c|c|c|c|@{}}
\multicolumn{1}{r}{\color{B}\fontfamily{ugq}\selectfont compression}
&	\TabTitle{sequential}
&	\TabTitle{\begin{minipage}{4em}coarse-\\[-.5ex]grained\end{minipage}}	
&	\TabTitle{\begin{minipage}{4em}fine-\\[-.5ex]grained\end{minipage}}	
&	\TabTitle{atomic}
\\
\cline{2-5}
\scshape dual-quantization	&			&			&	\YES	&			\\
\cline{2-5}
histogram					&			&			&	\YES	&	\YES	\\
\cline{2-5}
build Huffman tree			&	\YES	&			&	    	&			\\
\cline{2-5}
canonize codebook		&	\YES	&			&	\YES	&	\YES	\\
\cline{2-5}
Huffman encode (fix-length)	&			&       	&	\YES	&			\\
\cline{2-5}
deflate	(fix- to variable-length)&		&	\YES	&			&			\\	
\cline{2-5}
\multicolumn{1}{r}{\color{B}\fontfamily{ugq}\selectfont decompression}							\\
\cline{2-5}
inflate (Huffman decode)	&			&	\YES	&			&			\\	
\cline{2-5}
reversed \scshape dual-quantization&	&	\YES	&			&			\\
\cline{2-5}
\end{tabular}
    \caption{Parallelism implemented for \textsc{cuSZ}'s subprocedures (kernels) in compression and decompression.}
    \vspace{-6mm}
    \label{tab:how-to-par}
\end{table}

\subsection{Parallelizing Prediction-Quantization in Compression}
\label{sub:par-pred-quant}
In this section, we discuss our proposed optimization techniques to parallelize SZ's {\termPQ} procedure on GPU architectures. We first chunk the original data to gain coarse-grained parallelization, and then we assign a thread to each data point for fine-grained in-chunk parallel computations. 

\subsubsection{Chunking and Padding}
\label{subs:brute-force-par}

\begin{figure}[ht]
    \includegraphics[width=\linewidth]{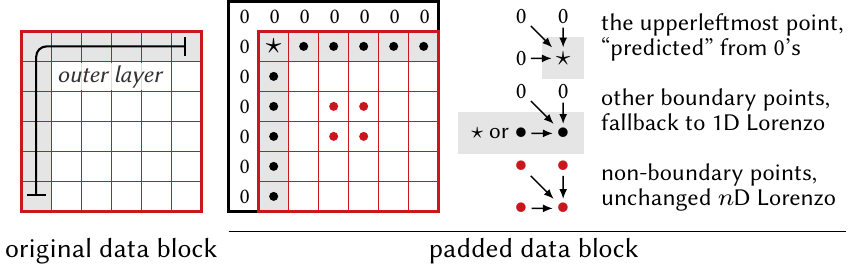}
    \caption{Data chunking and padding in \textsc{cuSZ}.}
    \vspace{-2mm}
    \label{fig:chunk-padding}
\end{figure}

\FIG~\ref{fig:chunk-padding} illustrates our chunking and padding technique.
For each chunked data block, we assign a thread to each data point (i.e., fine-grained parallelism). 
To avoid complex modifications to the prediction function after chunking, we add a padding layer to each block in 
the {\termPQ} step. 
We set all the values in the padding layer to \texttt{0}
such that they do not affect the predicted values of the points neighboring to the padding layer, as shown in \FIG~\ref{fig:chunk-padding}.
We note that in the original SZ, the uppermost points and leftmost points (denoted by ``outer layer'', shaded in \FIG~\ref{fig:chunk-padding}) are saved as unpredictable data directly.
In our chunking version, however, directly storing these points for each block would significantly degrade the compression ratio. Therefore, we apply $\ell$-prediction to the outer layer instead, such that every point in the block is consistently processed based on the $\ell$-predictor, avoiding thread/warp divergence.
Moreover, we initialize the padding layer with \texttt{0}s; the prediction for each outer-layer point falls back to 1D 1-order Lorenzo, as shown in \FIG~\ref{fig:chunk-padding}. Based on our empirical result, we adopt $32$ for 1D data, 16$\times$16 for 2D data, and 8$\times$8$\times$8 for 3D data.

\subsubsection{\textsc{Dual-Quantization} Scheme}
\label{subs:dual-quant-detail}
In the following discussion, we use circle $\circ$ and bullet $\bullet$ to denote the compression and decompression procedure, respectively.  We use star~$\star$ to denote all the values related to the data reconstruction in compression. The subscript $(\cdot)_{k}$ represents the $k$th iteration.

\begin{figure}[ht]
\centering
\begin{tikzpicture}
	\node[draw, rounded corners=12, text width=3.0in, align=center] {
	\begin{tabular}{@{}c@{}}
 		\includegraphics[width=2.8in]{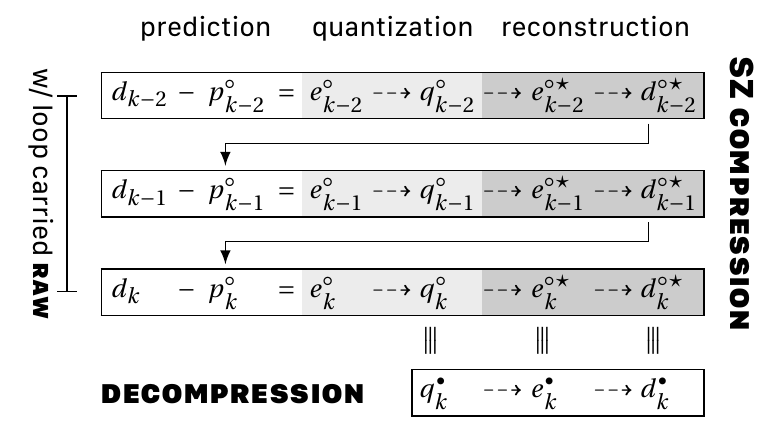}\\
 		\includegraphics[width=2.8in]{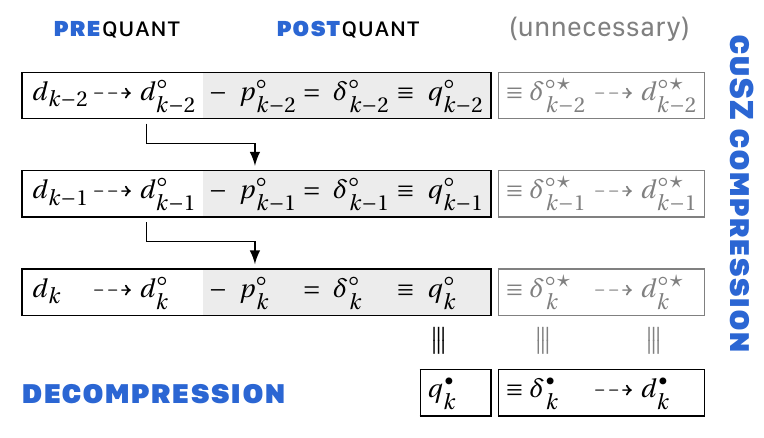}
	\end{tabular}
		};
	\draw (-1.54in, .0) -- ++(3.08in, 0);
\end{tikzpicture}
\caption{Diagram of original quanti\-za\-tion (\textit{top}) and \textsc{dual-quanti\-za\-tion} (\textit{bottom}) procedures. Arrow means data dependency.}
\vspace{-3mm}
\label{fig:graphic-cmp}

\end{figure}

\paragraph{Read-After-Write in SZ}
\label{par:raw}
In the original SZ algorithm, all data points need to go through {\termPQ}, and in situ reconstruction iteratively, which causes intrinsic \emph{read-after-write} (RAW) dependencies (as illustrated in \FIG~\ref{fig:graphic-cmp}). 

We describe the loop-carried RAW dependency issue in detail below.
For any data point at the $(k-1)$th iteration in SZ compression, given a predicted value $p_{k-1}$, the prediction error $e^{\circ}_{k-1}$ (i.e., $d_{k-1}^{\phantom{\circ}}\! - p^\circ_{k-1}$) is converted to an integer and a corresponding quantization code $q_{k-1}$ based on the user set error bound $eb$. 
Then, the reconstructed prediction error $e^{\circ\star}_{k-1}$ and the reconstructed value $d^{\circ\star}_{k-1}$
are generated by using $q_{k-1}$, $eb$, and $p_{k-1}^\circ$.
After that, $d^{\circ\star}_{k-1}$ is written back to replace $d_{k-1}$.
This procedure ensures that  $d^{\circ\star}_{k-1}$ is equivalent to the reconstructed $d^\bullet_{k-1}$ during decompression (as shown in \FIG~\ref{fig:graphic-cmp}); however, the $k$th iteration must wait until the update completes at the end of the $(k\!-\!1)$th iteration, which incurs loop-carried data dependency. Also note that $d^{\circ\star}$ is written back in the last step of the current iteration, and its written value is used at the beginning of the next iteration, therefore, the two consecutive iterations cannot overlap.
Hence, under the original design of the {\termPQ} in SZ, it is infeasible to effectively exploit fine-grained parallelism and efficiently utilize SIMT on GPUs. We present the original SZ's {\termPQ} step in Algorithm~\ref{algo:szcx} in detail.

\begin{algorithm}[ht]
\centering
    \caption{Original SZ of {\termPQ}}\label{algo:szcx}

\small\sffamily
\begin{algorithmic}[1] 
\For{$d\in D$}
    \Comment{{\color{B}\fontfamily{ugq}\selectfont compression}}
    \State $p^\circ \gets\ell(d_\textsc{\scshape sr})$, $e^\circ \gets p^\circ\! - d$
    \If{$e^\circ/eb < $ \textsc{cap (in-cap)} }   \Comment{quantization}
        \State $e^\circ_\text{D} \gets$ \textsc{integerize}$(e^\circ/(2\!\times\!eb))$
        \State \textsc{rehearsal} $\gets p^\circ + 2\cdot e^\circ_\text{D}\cdot eb$
        \State \textsc{watchdog}($\text{\scshape rehearsal} - d < eb$, fallback: \textsc{outlier})
    \Else
        \State \textsc{outlier: }$e^\circ_\text{D} \gets 0$ and record the verbatim $x\gets d$
    \EndIf
    \State $d\gets$ \textsc{rehearsal} or $x$ accordingly
    \Comment{incurs \textsc{\bfseries raw}}
\EndFor
\Statex
\For{$d^\bullet\in D^\bullet$ to reconstruct cascadingly}
\Comment{{\color{B}\fontfamily{ugq}\selectfont decompression}}
    \State $p^\bullet\gets\ell(d^\bullet_\text{\scshape sr})$
    \State $d^\bullet\gets p^\bullet + 2\cdot e^\circ_\text{D}\cdot eb$ if \textsc{in-cap} else verbatim $x$
\EndFor
\end{algorithmic}
\end{algorithm}

\paragraph{Proposed Dual-Quantization Approach}\label{par:dualquant}

To eliminate the RAW dependency, we propose a \textsc{dual-quanti\-za\-tion} scheme by modifying the data representation during the {\termPQ} procedure. 
Our \textsc{dual-quanti\-za\-tion} (abbreviated as \textsc{dual-quant}) consists of two steps: \textsc{pre\-quanti\-za\-tion} and \textsc{post\-quanti\-za\-tion}.

Given a dataset $D$ with an arbitrary dimension, we first quantize it based on the user-set $eb$ and convert it to a new dataset
\[
    D^\circ\! = \big\{d^\circ\!\mid d^\circ\! = \operatorname{round}\big\langle d / (2\times eb)\big\rangle,\ d\!\in D  \big\},
\]
where any $d^\circ \in D^\circ$ is strictly a multiple of $(2 \times eb)$. 
We call this step \textsc{prequantization} (abbreviated as \termPREQ). 
In order to avoid overflow, $d^\circ$ is stored in floating-point data type.
We note that the error introduced by {\termPREQ}
(defined as \textsc{posterror}) is strictly bounded by the user-set error bound, that is, $| d-2\cdot d^\circ\!\cdot eb| < eb$.
 
After the \textsc{prequantization}, we can calculate each predicted value based on its surrounding values (denoted by $d^\circ_\text{\scshape sr}$) and the $\ell$-predictor as
\[
    P^\circ\! = \{ p^\circ\! \mid p^\circ\! = \ell(d^\circ_\text{\scshape sr}),\  d^\circ\!\in D^\circ \}.
\]

The second step, called \textsc{postquantization} (abbreviated as \textsc{post\-quant}), serves as the counterpart of the linear-scaling quantization in the original SZ. 
{\termPOSTQ} computes the \emph{difference} between the predicted value and the \textsc{prequant}-ized value. Different from the original SZ, such difference does not cause any compression error (will be discussed later), we use $\delta$ instead of $e$ to denote this difference:
\[
    \Delta^\circ\!=\{\delta^\circ\! \mid \delta^\circ\! = d^\circ\! - p^\circ,\ d^\circ\!\in D^\circ,\ p^\circ\!\in P^\circ\}.
\]
Then, the quantization code $q^\circ$ is generated based on $\delta$.
Note that $q^\circ$ is quantitatively equivalent to $\delta^\circ$, represented differently: $\delta^\circ$ is a floating-point number to avoid subnormal values (i.e. under/overflow), while $q^\circ$ is an integer, which is used in the subsequent lossless coding (e.g., Huffman coding). 
It is worth noting that, during decompression, $d^\circ$ can be reconstructed (as $d^\bullet$) based on losslessly decoded $q^\bullet \equiv q^\circ$ (hence exactly $\delta^\circ$) and predicted $p^\circ$, thus this {\termPOSTQ}
step does not introduce any further error.

\paragraph{Eliminating RAW} In the following text, we explain in detail why the {\termDQ} method effectively eliminates the RAW dependency.
Conceptually, similar to the original SZ, we can construct $\delta^{\circ\star}$ and $d^{\circ\star}$ during the compression, as shown in \FIG~\ref{fig:graphic-cmp}.
In fact, for $(k\!-\!1)$th iteration, $\delta^{\circ\star}_{k-1}$ is strictly equal to $\delta^{\circ}_{k-1}$, because casting quantization code $q^\circ_{k-1}$ to $\delta^{\circ\star}_{k-1}$ is a exact reversed procedure of casting $\delta^{\circ}_{k-1}$ to $q^\circ_{k-1}$.

Similarly, $d^{\circ\star}_{k-1}$ and $d^\circ_{k-1}$ are also strictly equivalent.
Consequently, unlike the original SZ that must write $d^{\circ\star}_{k-1}$ back to update $d^{\circ\star}_{k-1}$ before the $k$th iteration, $d^{\circ\star}_{k-1} \equiv d^{\circ}_{k-1}$ always holds in our proposed {\termDQ} approach. 
As illustrated in \FIG~\ref{fig:graphic-cmp}, after {\termPREQ}, all $d^\circ$ are dependency free for \termPOSTQ.
By eliminating the loop-carried RAW dependency (marked as arrows in \FIG~\ref{fig:graphic-cmp}), we can effectively parallelize the {\termDQ} procedure by performing fine-grained (per-point) parallel computation, which is commonly seen in image processing~\cite{zhang2010image}.
We illustrate the detailed {\termDQ} procedure in Algorithm~\ref{algo:dq-szcx}. 

\begin{algorithm}[ht]
\centering
    \caption{\textsc{cuSZ} of {\termDQ}}\label{algo:dq-szcx}

\small\sffamily
\begin{algorithmic}[1] 
\For{$\forall d\in D$ concurrently}
    \Comment{{\color{B}\fontfamily{ugq}\selectfont compression}}
    \State \makebox[1.25em][l]{$d^\circ$}%
        \makebox[6em][l]{$\gets d/(2\!\times\!eb)$}%
        \Comment{(FP representation) \textsc{\textbf{pre}quant}} 
    \State \makebox[1.25em][l]{$d$}%
        \makebox[6em][l]{$\gets d^\circ$}%
        \Comment{{\sffamily\bfseries\scshape barrier}}
    \State \makebox[1.25em][l]{$p^\circ$}$\gets\ell(d^\circ_\text{\scshape sr})$, $\delta^\circ \gets p^\circ\! - d^\circ$
    \If{$\delta^\circ < $ \textsc{cap/2 (in-cap)} }   \Comment{\textsc{\textbf{post}quant}}
        \State $\delta^\circ_\text{D} \gets$ \textsc{cast\textless float2int\textgreater}$(\delta^\circ)$ 
    \Else
        \State \textsc{outlier: }$\delta^\circ_\text{D} \gets 0$ and record the verbatim $x\gets d^\circ$
    \EndIf
\EndFor
\Statex
\For{$d^\bullet\in D^\bullet$ to reconstruct cascadingly}
\Comment{{\color{B}\fontfamily{ugq}\selectfont decompression}}
    \State $p^\bullet\gets\ell(d^\bullet_\text{\scshape sr})$
    \State $d^\bullet\gets (p^\bullet + \delta^\circ_\text{D})\cdot (2\!\times\!eb)$ if \textsc{in-cap} else verbatim $x$
\EndFor
\end{algorithmic}
\end{algorithm}

\paragraph{Lorenzo Predictor with Binomial Coefficients}\label{par:lorenzo-int-op}

According to Tao et al.~\cite{sz17}, the generalized $\ell$-predictor is given by
\[
\begin{split}
   \textstyle
	\sum
		^{{k_{1\ldots d}\neq\mathbf{0}}}
		_{{0\le k_{1\ldots m} \le n}}
   \left\langle\prod^m_{j=1}(-1)^{k_j+1} {n \choose k_j} \right\rangle \cdot d_{x_1 - k_1,\cdots, x_d - k_d} ,
\end{split}
\]
where
$
	\textstyle \sum
		^{{k_{1\ldots d}\neq\mathbf{0}}}
		_{{0\le k_{1\ldots m} \le n}}
   \left\langle\prod^m_{j=1}(-1)^{k_j+1} {n \choose k_j} \right\rangle = 1
$ and $d\in D$.
For example, 1D 1-order $\ell$-predictor is $p^\circ_a = d^\circ_{a-1}$, and 2D 1-order $\ell$-predictor is $p^\circ_{(a,b)}=\ell\left(d^\circ_\text{sr}\right)=d^\circ_{a-1,b} + d^\circ_{a,b-1} - d^\circ_{a-1,b-1}$, as illustrated in \FIG~\ref{fig:overview}.
We note that all the coefficients in the formula of the $\ell$-predictor are integers; thus, the prediction computation consists of mathematically integer-based operations (additions and multiplications) and results in unit weight. This ensures that no division is needed, and the data reconstruction based on {\termDQ} is fairly precise and robust with respect to machine $\epsilon$, however, the original SZ using precise floating-point operations suffers from underflow. Note that the predicted values which are integers will be completely corrected by the saved quantization codes in decompression, so the final error is still bounded by $eb$.

\subsection{Efficient Customized Huffman Coding}
\label{sub:huffman}

To efficiently compress the quantization codes generated by {\termDQ}, we develop an efficient customized Huffman coding for SZ on GPUs. 
Specifically, Huffman coding consists of the following subprocedures:
\Circled{1} calculate the statistical frequency for each quantization bin (as a symbol);
\Circled{2} build the Huffman tree based on the frequencies and generate a base codebook along with each code bitwidth; 
\Circled{3} transform the base codebook to the canonical Huffman codebook (called canonization);
\Circled{4} encode in parallel according to the codebook, and concatenate Huffman codes into a bitstream (called \textit{deflating}).
And Huffman decoding is composed of 
\Circled{1} retrieving the reverse codebook and 
\Circled{2} decoding accordingly.

Note that the fourth subprocedure of encoding can be further decomposed into two steps for fine-grained optimization. 
Codebook-based encoding is basically memory copy and can be fully parallelized in a fine granularity, whereas deflating can be  performed only sequentially (except blockwise parallelization discussed in \SEC\ref{subs:brute-force-par}) because of its atomic operations.
We discuss Huffman coding on GPUs step by step as follows.

\subsubsection{Histogram for Quantization Bins}
The first step of Huffman coding is to build a histogram representing the frequency of each quantization bin from the data prediction step. 
The GPU histogramming algorithm that we use is derived from the algorithm proposed by G\'omez-Luna et al.~\cite{GmezLuna2012AnOA}. This algorithm minimizes conflicts in updating the histogram bin locations by replicating the histogram for each thread block and storing the histogram in shared memory. Where possible, conflict is further reduced by replicating the histogram such that each block has access to multiple copies. All threads inside a block read a specified partition of the quantization codes and use atomic operations to update a specific replicated histogram. As each block finishes its portion of the predicted data, the replicated histograms are combined via a parallel reduction into a single global histogram, which is used to construct the final codebook in Huffman coding.

\subsubsection{Constructing Huffman Codebook}
\label{subs:build-codebook}

In order to build the optimal Huffman tree, the local symbol frequencies need to be aggregated to generate the global symbol frequencies for the whole dataset. By utilizing the aggregated frequencies, we build a codebook according to the Huffman tree for encoding.
Note that the number of symbols---namely, the number of quantization bins---is a limited number (generally no greater than 65,536) that is much smaller than the data size (generally, millions of data points or more). This leads to a much lower number of nodes in the Huffman tree compared with the data size, such that the time complexity of building a Huffman tree is considered low. 
We note that building Huffman tree sequentially on CPU benefits from high CPU-frequency and low memory-access latency. However, it requires CPU-to-GPU/GPU-to-CPU transfer of frequencies/codebook before/after building the tree, and communicating these two small messages would incur non-negligible overheads. Therefore, we adopt one GPU thread to build the Huffman tree sequentially to avoid such overheads.

\begin{figure}[!htbp]
    \centering
    \includegraphics[]{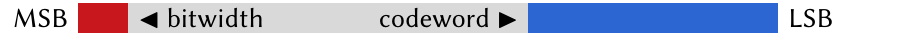}
    \caption{Fixed-length representation of Huffman codeword and its bitwidth.}
    \label{fig:fixed-len}
    \vspace{-2mm}
\end{figure}

We propose an adaptive codeword representation to enhance the utilization of memory bandwidth, which improves the Huffman encoding performance in turn. We illustrate the organization of the codebook in \FIG~\ref{fig:fixed-len}.
The codebook is organized by units of unsigned integers, and each contains a variable-length Huffman codeword from LSB (the rightmost or the least significant bits) and its bitwidth from MSB (the leftmost or the most significant bits).
According to the pessimistic estimation of maximum bitwidth of optimal Huffman codeword~\cite{abu2000maximal}, one is supposed to use \verb+uint64_t+ to hold each bitwidth-codeword representation. 
For example, the maximum bitwidth could be 33 bits for \texttt{CLDHGH} field from \texttt{CESM-ATM} dataset in the worst case. 
However, we note that using 
\verb+uint32_t+ 
to represent a bitwidth-codeword tuple can significantly improve the Huffman coding and decoding performance compared with using 64-bit unsigned integers (i.e., \verb+uint64_t+), because of higher GPU memory bandwidth utilization.
Thus, we propose to dynamically select \verb+uint32_t+ or \verb+uint64_t+ representation for the Huffman bitwidth-codeword based on the practical maximum bitwidth instead of pessimistic estimation. We show the performance evaluation with different representations in \SEC\ref{sec:evaluation}.

The theoretical time complexity is $\mathcal{O}(k\log k)$ for building a Huffman tree and $\mathcal{O}(k)$ for a traversing tree,
where $k$ is the number of symbols (quantization bins). 
Our experiments show that the real execution time of building a Huffman tree is consistent with the theoretical time complexity analysis (see \TAB~\ref{tab:build-codebook}).
On the other hand, the number of symbols is determined by the smoothness of the dataset and the user-desired error bound (1,024 by default). For example, with a relatively large error bound such as the value-range-based relative error bound \footnote{Value-range-based relative error bound (denoted by \texttt{valrel}) is the error bound relative to the value range of the dataset.} of $10^{-3}$, we observe that most of the symbols are concentratedly distributed near the central of codebook. As the error bound decreases, the symbols become more evenly distributed. Thus, determining a suitable number of quantization bins is important for high performance in constructing a codebook.

\subsubsection{Canonizing Codebook}

A canonical Huffman codebook~\cite{schwartz1964generating} holds the same bitwidth of each codeword as the original Huffman codebook (i.e., base codebook), while its bijective mapping (between quantization code and Huffman codeword) and variable codeword make the memory layout organized more efficiently.
The time complexity of sequentially canonizing codebook from the base is $\mathcal{O}(k)$, where $k$ is the number of symbols (i.e., the number of quantization bins) and is sufficiently small compared with the data size. 
By using a canonical codebook, we can (i) decode without the Huffman tree, (ii) efficiently cache the reverse codebook for high decoding throughput, and (iii) maintain exactly the same compression ratio as the base Huffman codebook.

The process of canonizing codebook can be decomposed into the following subprocedures:
\Circled{1} linear scanning of the base codebook (sequentially $\mathcal{O}(k)$), which is parallelized at fine granularity with atomic operations; 
\Circled{2} loosely radix-sorting of the codewords by bitwidth (sequentially $\mathcal{O}(k)$), which cannot be parallelized because of the intrinsic RAW dependency; and
\Circled{3} building the reverse codebook (sequentially $\mathcal{O}(k)$), which is enabled with fine-grained parallelism.

It is intuitive to separate the functionalities of the aforementioned subprocedures and implement them into independent CUDA kernels with different configurations 
(i.e., \texttt{blockDim} and \texttt{gridDim}).
Based on our profiling results on NVIDIA V100 GPU, however, launching a CUDA kernel usually takes about 60 microseconds ($\mu$s)  (about 200 $\mu$s for the three kernels of canonization) measured by 11 kernels launched in total. 
Moreover, any two consecutive subprocedures require an additional expensive synchronization (i.e., \verb|cudaDeviceSynchronize|).
However, our experiment indicates that canonizing codebook is sufficiently fast; thus, we integrate all the three subprocedures in one single kernel. 

We note that this single kernel must be launched with more threads than the that is limited for a single thread block (i.e. 1024) for two reasons. On the one hand, a high scalability is required for the parallel reads/writes to match the problem size in subprocedures \Circled{1} and \Circled{3}. On the other hand, unlike histogramming that saves only the $\Theta(k)$ frequencies in shared memory, this kernel requires saving both the codebook and its footprint, which may exceed the maximum allowable capacity of shared memory in a single thread block (e.g., 96 KB for a V100). 
Since shared memory is only visible to its designated thread block, shuffling codewords in shared memory across different thread blocks is semantically prohibitive. In addition, there is few intermediate data reuse, thus, we use global memory instead of shared memory to save the codebook.
Hence, we employ the state-of-the-art CUDA API---Cooperative Groups~\cite{harris2017cooperative}---to achieve in-grid operation. Specifically, we launch the same number of threads as the codeword in the base codebook. We select one thread to perform the RAW-restricted sequential subprocedure when needed (see \TAB~\ref{tab:how-to-par}). Note that it takes only 32 $\mu$s on a V100 to launch this Cooperative Groups enabled kernel, which significantly reduces the overhead compared to launching multiple kernels. 
Moreover, compared to two inter-kernel barriers with more than 2$\times$60~$\mu$s, two in-grid barriers have relatively low overheads, and eventually result in, for instance, about 200 $\mu$s kernel time with $k=1024$.

\subsubsection{Encoding and Deflating}
\label{subs:huff-coding}

We design an efficient solution to perform the encoding by GPU threads in parallel. Encoding involves looking up a symbol in the codebook and performing a memory copy.
After we adaptively select a 32-/64-bit unsigned integer to represent a Huffman code with its bitwidth, the encoding step is 
\textit{massively} parallelized.
To generate the \emph{dense} bitstream of Huffman codes within each data block, we conduct \emph{deflating} in order to concatenate the Huffman codes and remove the unnecessary zero bits according to the saved bitwidths.

Since the deflated code is organized sequentially, we apply the coarse-grained chunkwise parallelization technique discussed in \SEC\ref{subs:brute-force-par}. In particular, a data chunk for compression and decompression is mapped to one GPU thread.
Note that the chunk size for deflating is not necessarily the same as 
the chunk size for {\termDQ}, 
and it does not rely on the dimensionality.
We optimize the deflating chunk size by evaluating the performance with different sizes (will be showed in \SEC\ref{subs:eval-cx}). 
We also employ memory reuse technique to reduce the GPU memory footprint in deflating. Specifically, we reuse the memory space of Huffman codes for the deflated bitstream because the latter uses significantly less memory space and does not have any conflict when writing the deflated bitstream to the designated location.

\subsection{Decompression}
\label{sec:decompression}

\textsc{cuSZ}'s decompression consists of two steps: Huffman decoding (or inflating the densely concatenated Huffman bitstream) and reversed {\termDQ}. In inflating, we first use the previously built reverse codebook to retrieve the quantization codes from the deflated Huffman bitstream. Then, based on the retrieved quantization codes, we reconstruct the floating-point data values. Note that only coarse-grained chunking can be applied to decompression, and its chunk size is determined in compression.  
The reason is that the two steps both have a RAW dependency issue. In fact, retrieving the variable-length codes has the same pattern as loop-carried RAW dependency. For the reversed dual-quantization procedure, each data point cannot be decompressed until its preceding values are fully reconstructed. 

\section{Experimental Evaluation}
\label{sec:evaluation}

In this section, we present our experimental setup (including platform, baselines, and datasets) and our evaluation results.

\subsection{Experimental Setup}
\label{sub:evalsetup}

\paragraph{Evaluation Platform}
We conduct our experimental evaluation using PantaRhei cluster \cite{pantarhei}.
We perform the experiments on an NVIDIA V100 GPU~\cite{nv100} from the cluster and compare with lossy compressors on two 20-core Intel Xeon Gold 6148 CPUs from the cluster. The GPU is connected to the host via 16-lane PCIe 3.0 interconnect. We use NVIDIA CUDA 9.2 and its default profiler to measure the kernel time.

\begin{table}
    \footnotesize\centering
    \begin{tabular}{@{} l r r r @{}}
\tableTitleFormat datasets
    & \tableTitleFormat type
    & \datumSize{\footnotesize datum size}{\tableTitleFormat dimensions}
    & \dataTableBase{\#fields}{\tableTitleFormat example(s)}{r}
\\ [.5ex]
\toprule
\datasetName{cosmology}{HACC}
    & \ttfamily fp32
    & \datumSize{1,071.75 MB}{280,953,867}
    & \datasetField{6}{x, vx}
\\
\datasetName{climate}{CESM-ATM}
    & \ttfamily fp32
    & \datumSize{24.72 MB}{1,800$\times$3,600} 
    & \datasetField{79}{CLDHGH, CLDLOW}
\\
\datasetName{climate}{Hurricane}
    & \ttfamily fp32
    & \datumSize{95.37 MB}{100$\times$500$\times$500} 
    & \datasetField{20}{CLOUDf48, Uf48}
\\
\datasetName{cosmology}{Nyx}
    & \ttfamily fp32
    & \datumSize{512.00 MB}{512$\times$512$\times$512}
    & \datasetField{6}{baryon\_density}
\\
\datasetName{quantum}{QMCPACK}
    & \ttfamily fp32
    & \datumSize{601.52 MB}{288$\times$115$\times$69$\times$69}
    & \datasetField{2 formats}{einspline}
\\
\bottomrule
\end{tabular}
    \caption{Real-world datasets used in evaluation.}
    \vspace{-4mm}
    \label{tab:datasets}
\end{table}

\paragraph{Comparison Baselines}
We compare our \textsc{cuSZ} with two baselines: SZ-1.4.13.5 and cuZFP \cite{cuZFP}. For SZ-1.4, we adopt the default
setting: 16 bits for linear-scaling quantization (i.e., 1,024 quantization bins), \texttt{best\_compression} mode, and
\texttt{best\_speed} mode for \texttt{gzip}, which lead to a good tradeoff between compression ratio and performance.

\begin{table}
\centering\footnotesize
\sffamily
\begin{tabular}{@{}l 
    *{4}{>{\raggedleft\arraybackslash}p{2.0em}} 
    *{3}{>{\raggedleft\arraybackslash}p{2.2em}} 
@{}}
    \singleLineTableHeadFormat\#quant. 
    & \singleLineTableHeadFormat 128 
    & \singleLineTableHeadFormat 256
    & \singleLineTableHeadFormat 512
    & \singleLineTableHeadFormat 1024
    & \singleLineTableHeadFormat 2048 
    & \singleLineTableHeadFormat 4096 
    & \singleLineTableHeadFormat 8192 
    \\
\toprule
\sffamily build tree	
    &	{0.48}
    &	0.77
    &   1.80
	&	{2.13}
    &   6.46
    &   12.68
    &   25.06
\\
\sffamily get codebook
    &   {0.20}
    &   1.14
    &   2.36
	&	{2.69}
    &   7.09
    &   14.43
    &   25.65
\\
\sffamily total
    &   {0.68}
    &   2.16
    &   4.16
	&	{4.81}
    &   13.55
    &   27.10
    &   50.71
\\
\bottomrule
\end{tabular}
\caption{Breakdown time (in ms) of constructing a codebook, including building a Huffman tree and creating a codebook according to the tree based on the Hurricane Isabel dataset.}
\vspace{-4mm}
\label{tab:build-codebook}
\end{table}

\paragraph{Test Datasets}
We conduct our evaluation and comparison based on five typical real-world HPC simulation datasets of each dimensionality from the Scientific Data Reduction Benchmarks suite~\cite{sdrbench}: \Circled{1} 1D \texttt{HACC} cosmology particle simulation~\cite{hacc}, \Circled{2} 2D \texttt{CESM-ATM} climate simulation~\cite{cesm-atm}, \Circled{3} 3D \texttt{Hurricane ISABEL} simulation~\cite{hurricane}, \Circled{4} 3D \texttt{Nyx} cosmology simulation~\cite{nyx}, and \Circled{5} 4D \texttt{QMCPACK} quantum Monte Carlo simulation~\cite{qmcpack}. 
They have been widely used in prior works~\cite{tao2018optimizing,liang2018error,xincluster18,use-case-Franck,liang2019improving} and are good representatives of production-level simulation datasets.
\TAB~\ref{tab:datasets} shows all 112 fields \footnote{The \texttt{QMCPACK} dataset includes only one field but with two representations.} across these datasets. The data sizes for the  five datasets are 6.3 GB, 2.0 GB, 1.9 GB, 3.0 GB, and 1.2 GB, respectively. 
Note that our evaluated HACC dataset is consistent with real-world scenarios that generate petabytes of data. For example, according to ~\cite{hacc}, a typical large-scale HACC simulation for cosmological surveys runs on 16,384 nodes each with 128 million particles and generates 5 PB over the whole simulation. The simulation contains 100 individual snapshots of roughly 3 GB per node.
We evaluate a single snapshot for each dataset instead of all the snapshots, because the compressibility of most of the snapshots usually has strong similarity.
Moreover, when the field is too large to fit in a single GPU's memory, \textsc{cuSZ} divides it into blocks and then compresses them block by block.

\subsection{Evaluation Results and Analysis}
\label{sub:evals}
In this section, we evaluate the compression performance and quality of \textsc{cuSZ} and compare it with CPU-SZ and cuZFP.

\subsubsection{Compression Performance}
\label{subs:eval-cx}
We first evaluate the performance of {\termDQ} of \textsc{cuSZ}. The average throughput of the {\termDQ} step on each tested dataset is shown in \TAB~\ref{tab:breakdown-kernel}. Compared with the original serial CPU-SZ, the {\termPQ} throughput is improved by more than 1000$\times$ via our proposed {\termDQ} on the GPU. This improvement is because {\termDQ} entirely eliminates the RAW dependency and leads to fine-grained (per-point) parallel computation, which is significantly accelerated on the GPU. 

We then evaluate the performance of our implemented Huffman coding step by step. First, we conduct the experiment of Huffman histogram computation and show its throughput performance\footnote{All throughputs shown are measured based on the original data size and time.}. Efficiently computing a histogram on a GPU is an open challenging problem, {because of the way that multiple threads need to write to the same memory locations simultaneously.}
Here, we present a method that, while a bottleneck in the Huffman process, is a 2$\times$ improvement from a serial implementation.

Next, we perform the experiment of constructing codebook with different numbers of quantization bins, as shown in \TAB~\ref{tab:build-codebook}. We note that the execution times of building a Huffman tree and creating a codebook are consistent with our time complexity analyses in \SEC\ref{subs:build-codebook}. We use 1,024 quantization bins by default. Since the time overhead of constructing a codebook  depends only on the number of quantization bins, it is almost fixed---for example, \behold{4.81} ms---for the remaining experiments. We also note that a larger data size lowers the relative performance overhead of constructing a codebook, thus leading to higher overall performance.

\begin{table}[ht]
\footnotesize\sffamily
\centering
\sffamily
\begin{tabular}{@{}l@{}r@{}rrrrr@{}}
&
& \doubleLineTableHead{r@{}}{\scriptsize\color{GRAY}1071 MB}{hacc}
& \doubleLineTableHead{r@{}}{\scriptsize\color{GRAY}25 MB}{cesm-atm}
& \doubleLineTableHead{r@{}}{\scriptsize\color{GRAY}95 MB}{hurricane}
& \doubleLineTableHead{r@{}}{\scriptsize\color{GRAY}512 MB}{nyx} 
& \doubleLineTableHead{r@{}}{\scriptsize\color{GRAY}602 MB}{qmcpack}
\\
\toprule
   enc.64 
&   \scriptsize\color{GRAY} $\mu$s
&   \scriptsize\color{GRAY} 4,274.3 
&   \scriptsize\color{GRAY} 97.1
&   \scriptsize\color{GRAY} 385.8 
&   \scriptsize\color{GRAY} 2,044.7 
&   \scriptsize\color{GRAY} 2,401.4 
\\[-.5ex]
&   \scriptsize GB/s
&   250.9
&   255.1
&   251.7
&   251.1
&   251.1
\\
\midrule
   enc.32
&  \scriptsize\color{GRAY}   $\mu$s
&  \scriptsize\color{GRAY}   2,839.3
&  \scriptsize\color{GRAY}   64.1
&  \scriptsize\color{GRAY}   255.8 
&  \scriptsize\color{GRAY}   1,358.6   
&  \scriptsize\color{GRAY}   1,595.6
\\[-.5ex]
&   \scriptsize GB/s
&	377.7
&	386.6
&	379.6
&	377.9
&	377.9 
\\
\bottomrule
\end{tabular}

\caption{Performance of encoding and deflating based on the constructed codebook (averaged based on all fields for each set).}
\vspace{-2mm}
\label{tab:enc-deflate}
\end{table}

We also evaluate the performance of encoding and decoding based on the canonical codebook. To increase the memory bandwidth utilization, we adapt online selection of Huffman codeword representation between a \verb+uint32_t+ and a \verb+uint64_t+.
\TAB~\ref{tab:enc-deflate} illustrates that our encoding achieves about \behold{250} GB/s for \verb+uint64_t+ and about \behold{380} GB/s \footnote{NVIDIA V100 GPU has a theoretical peak memory bandwidth of 900 GB/s.} for \verb|uint32_t|, based on the test with all 111 fields under the error bound of 1e-4.  
Hence, we conclude that using a uint32\_t enables significantly higher performance than using a uint64\_t. 
Because of the coarse-grained chunk-wise parallelization, the performance of deflating is about \behold{60} GB/s, which is lower than the encoding throughput of \behold{380} GB/s. 
Consequently, the Huffman coding performance is  bounded mainly by the deflating throughput.

To improve the deflating and inflating performance, we further evaluate different chunk sizes and identify the appropriate sizes for both deflating and inflating on the tested datasets, as shown in \TAB~\ref{tab:concurrency}.
Specifically, we evaluate  chunk sizes ranging from $2^6$ to $2^{16}$, due to different field sizes. We observe that using a total of around 2e4 concurrent threads consistently achieves the optimal throughput.
Note that inflating must follow exactly the same data chunking strategy as deflating; thus we need to select the same chunk size. 
Even under this constraint, our selected chunk sizes still achieve throughputs close to the peak ones, as illustrated in \TAB~\ref{tab:concurrency}. 
Therefore, we conclude that the overall optimal performance can be achieved by setting up a total of 2e4 concurrent threads in practice.

\begin{table}[ht]
    \centering\sffamily\footnotesize
    \renewcommand{\arraystretch}{1.15}
\begin{tabular}{@{}r|rrr|rrr|@{}}
    \multicolumn{1}{c}{}
&   bitrate &   CR  & \multicolumn{1}{c}{PSNR}
&   bitrate &   CR  & \multicolumn{1}{c}{PSNR}
\\
\cline{2-4}\cline{5-7}
\scshape cesm-atm  &   3.08 bits   &   10.4    &   85.3 dB     &   12 bits &   2.7    &   88.7 dB
\\
\scshape hurricane  &   3.45 bits   &   9.3    &   87.0 dB     &   12 bits &   2.7    &   81.9 dB
\\
\scshape nyx        &   2.49 bits   &   12.8   &   86.0 dB     &    6 bits &   5.3    &   85.1 dB
\\
\scshape qmcpack    &   3.38 bits   &   9.5    &   85.0 dB     &    8 bits &   4.0    &   84.0 dB
\\[.5ex]
\cline{2-4}\cline{5-7}
    \multicolumn{1}{c}{}
&   \multicolumn{3}{c}{\bfseries\textsc{cuSZ}}
&   \multicolumn{3}{c}{\bfseries cuZFP}
\\
\end{tabular}
    \caption{Bitrate comparison at PSNR of about 85 dB (\textsc{cuSZ}'s PSNRs are no lower than cuZFP's). CR stands for compression ratio.}
    \label{tab:rate-distortion}
    \vspace{-2mm}
\end{table}

\begin{figure}[ht]
    \includegraphics[width=\linewidth]{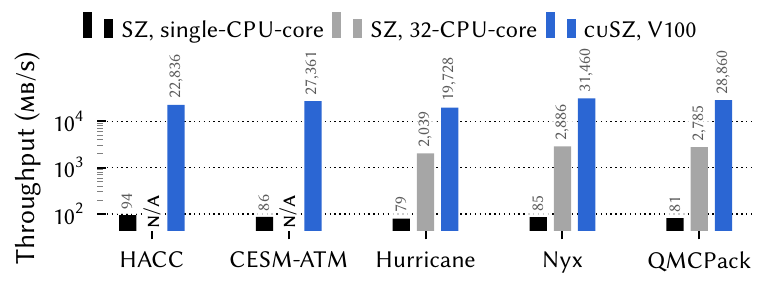}
    \includegraphics[width=\linewidth]{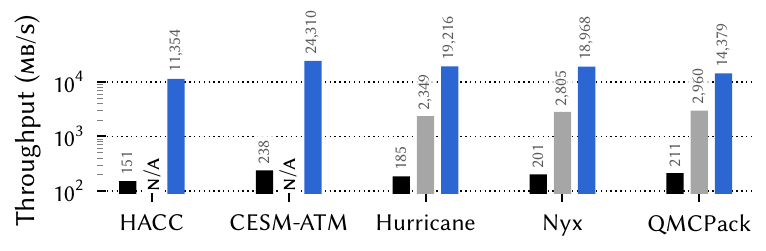}
    \caption{Compression (\textsc{top}) and decompression (\textsc{bottom}) throughput of \textsc{cuSZ} and CPU-SZ on tested datasets.}
    \vspace{-2mm}
    \label{fig:kernel-throughput}
\end{figure}

\begin{table*}[ht]
\footnotesize\sffamily
\newcommand{\CTITLE}{\scshape\sffamily\bfseries\small}
\newcommand{\HLNUM}[1]{\bfseries #1}

\resizebox{\linewidth}{!}{%
\renewcommand{\arraystretch}{1.05}
\begin{tabular}{@{}l@{}}
\\
\sffamily\scshape\bfseries chunk \\
\sffamily\scshape\bfseries size  \\
\toprule
2$^\text{6}$  \\
2$^\text{7}$  \\
2$^\text{8}$  \\
2$^\text{9}$  \\
2$^\text{10}$ \\
2$^\text{11}$ \\
2$^\text{12}$ \\
2$^\text{13}$ \\
2$^\text{14}$ \\
2$^\text{15}$ \\
2$^\text{16}$ \\
\bottomrule
\end{tabular}
\enspace

\begin{tabular}{@{}rrr@{}}
\multicolumn{3}{@{}l}{\CTITLE hacc}
\\
\multicolumn{3}{@{}l}{\scshape\sffamily\color{gray} 1071.8 mb\quad 280,953,867 \sf f32} 
\\
  \scshape\sffamily\#thread 
& \scshape\sffamily deflate 
& \scshape\sffamily inflate
\\
\toprule
       $.$    &    $.$    &    $.$     \\
       $.$    &    $.$    &    $.$     \\
       $.$    &    $.$    &    $.$     \\
       $.$    &    $.$    &    $.$     \\
       $.$    &    $.$    &    $.$     \\
    1.4e5 & 4.6   & 2.8 \\
    6.9e4 & 5.1   & 5.1 \\
    3.4e4 & 13.6  & 12.1 \\
    \HLNUM{1.7e4} & \HLNUM{63.1} & \HLNUM{35.0} \\
    8.6e3 & 65.8  & 28.1 \\
    4.3e3 & 45.9  & 14.3 \\
\bottomrule
\end{tabular}
\enspace

\begin{tabular}{@{}rrr@{}}
\multicolumn{3}{@{}l}{\CTITLE cesm}
\\
\multicolumn{3}{@{}l}{\scshape\sffamily\color{gray} 24.7 mb\quad 6,480,000 \sf f32} 
\\
  \scshape\sffamily\#thread 
& \scshape\sffamily deflate 
& \scshape\sffamily inflate
\\
\toprule
    {1.0e5} & 11.3 & 25.0 \\
    {5.1e4} & 15.5 & 37.8 \\
     \HLNUM{2.5e4}  &  \HLNUM{67.1}   &  \HLNUM{41.6}   \\
    {1.3e4} & 55.6 & 30.7 \\
    {6.3e3} & 48.2 & 19.6 \\
       $.$    &    $.$    &    $.$     \\
       $.$    &    $.$    &    $.$     \\
       $.$    &    $.$    &    $.$     \\
       $.$    &    $.$    &    $.$     \\
       $.$    &    $.$    &    $.$     \\
       $.$    &    $.$    &    $.$     \\
    \bottomrule
\end{tabular}
\enspace

\begin{tabular}{@{}rrr@{}}
\multicolumn{3}{@{}l}{\CTITLE hurricane}
\\
\multicolumn{3}{@{}l}{\scshape\sffamily\color{gray} 95.4 mb\quad 25,000,000 \sf f32} 
\\
  \scshape\sffamily\#thread 
& \scshape\sffamily deflate 
& \scshape\sffamily inflate
\\
\toprule
       $.$    &    $.$    &    $.$     \\
       $.$    &    $.$    &    $.$     \\
    {9.8e4} & 5.1 & 11.0 \\
    {4.9e4} & 10.2 & 9.4 \\
     \HLNUM{2.4e4}  &  \HLNUM{64.6}   &  \HLNUM{34.2}   \\
    {1.2e4} & 57.3 & 27.7 \\
    {6.1e3} & 50.7 & 17.8 \\
       $.$    &    $.$    &    $.$     \\
       $.$    &    $.$    &    $.$     \\
       $.$    &    $.$    &    $.$     \\
       $.$    &    $.$    &    $.$     \\
    \bottomrule
\end{tabular}
\enspace

\begin{tabular}{@{}rrr@{}}
\multicolumn{3}{@{}l}{\CTITLE nyx} 
\\
\multicolumn{3}{@{}l}{\scshape\sffamily\color{gray} 512 mb\quad 134,217,728 \sf f32} 
\\
  \scshape\sffamily\#thread 
& \scshape\sffamily deflate 
& \scshape\sffamily inflate
\\
\toprule
       $.$    &    $.$    &    $.$     \\
       $.$    &    $.$    &    $.$     \\
       $.$    &    $.$    &    $.$     \\
       $.$    &    $.$    &    $.$     \\
    {1.3e5} & 4.7 & 5.9 \\
    {6.6e4} & 5.7 & 6.3 \\
    {3.3e4} & 25.1 & 16.1 \\
     \HLNUM{1.6e4}  &  \HLNUM{69.7}   &  \HLNUM{52.4}   \\
    {8.2e3} & 72.4 & 42.6 \\
    {4.1e3} & 50.0 & 23.1 \\
       $.$    &    $.$    &    $.$     \\
    \bottomrule
\end{tabular}
\enspace

\begin{tabular}{@{}rrr@{}}
\multicolumn{3}{@{}l}{\CTITLE qmcpack}
\\
\multicolumn{3}{@{}l}{\scshape\sffamily\color{gray} 601.5 mb\quad 157,684,320 \sf f32}
\\
  \scshape\sffamily\#thread 
& \scshape\sffamily deflate 
& \scshape\sffamily inflate
\\
\toprule
       $.$    &    $.$    &    $.$     \\
       $.$    &    $.$    &    $.$     \\
       $.$    &    $.$    &    $.$     \\
       $.$    &    $.$    &    $.$     \\
    {1.5e5} & 4.7 & 5.1 \\
    {7.7e4} & 5.2 & 6.2 \\
    {3.8e4} & 12.9 & 11.1 \\
     \HLNUM{1.9e4}  &  \HLNUM{72.7}   & \HLNUM{40.3} \\
    {9.6e3} & 75.9 & 29.0 \\
    {4.8e3} & 56.0 & 16.1 \\
       $.$    &    $.$    &    $.$     \\
    \bottomrule
\end{tabular}%
}
\caption{Throughputs (in GB/s) versus different numbers of threads launched on V100. The optimal thread number in terms of inflating and deflating throughput is shown in bold.}
\label{tab:concurrency}
\vspace{-2mm}
\end{table*}

\begin{table*}[ht]
    \footnotesize\centering\sffamily
    \resizebox{\linewidth}{!}{
\begin{tabular}{@{} l@{} r@{} rrr rr rr @{}}   
&
& \grandTableTitle{r}{predict. (p)}{\bfseries + quant. (q)}
& 
& \grandTableTitle{r}{huffman}{}
& 
& \grandTableTitle{r}{kernel}{compression}
& \grandTableTitle{r}{gpu-to-cpu}{\footnotesize \textsc{valrel}@10\textsuperscript{-4}}
& \grandTableTitle{r}{overall}{compression}
   \\ 
\toprule    
&       
& \TabUnitStyle mb/s
&       
& \TabUnitStyle mb/s
&       
& 
&       
& \TabUnitStyle mb/s
\\
\textsc{\sffamily CPU-SZ} 
& \textsc{\sffamily hacc}  
& 137.7 
&       
& 328.6 
&       
& - 
& -     
& 94.1 \\
          
& \textsc{\sffamily cesm-atm} 
& 105.0
&       
& 459.1 
&       
& - 
& -     
& 85.5 \\
          
& \textsc{\sffamily hurricane} 
& 93.8 
&       
& 504.0
&       
& - 
& -     
& 78.5 \\
          
& \textsc{\sffamily nyx}   
& 98.5 
&       
& 648.7 
&       
& - 
& -     
& 84.7 \\
          
& \textsc{\sffamily qmcpack} 
& 97.5 
&       
& 396.2 
&       
& - 
& -     
& 80.8 \\
\midrule
&
& 
& {\color{black}\scshape histogram}
& {\color{black}\scshape codebook}
& {\color{black}\scshape coding}
& 
& 
& 
\\[-.4ex]
&       
& \TabUnitStyle gb/s
& \TabUnitStyle gb/s
& \TabUnitStyle ms
& \TabUnitStyle gb/s
& \TabUnitStyle gb/s
& \TabUnitStyle gb/s
& \TabUnitStyle gb/s
\\
\textsc{\sffamily\scshape cuSZ} 
& \textsc{\sffamily hacc}  
& 207.7
& 602.8 
& 5.16
& 54.1
& 40.0
& 53.2	
& 22.8 
\\          
& \textsc{\sffamily cesm-atm} 
& 252.1
& 345.3 
& 4.33
& 57.2
& 41.1 
& 81.9	
& 27.4 
\\          
& \textsc{\sffamily hurricane} 
& 175.8
& 418.0
& 4.81 
& 55.2 
& 38.2 
& 40.8	
& 19.7	
\\          
& \textsc{\sffamily nyx}   
& 200.2
& 427.6 
& 3.84
& 58.8
& 41.1 
& 134.1	
& 31.6 
\\
& \textsc{\sffamily qmcpack} 
& 189.6
& 346.1 
& 4.09
& 61.0
& 40.7 
& 99.2	
& 28.9 
\\
\midrule
{\sffamily cuZFP} 
& \textsc{\sffamily hacc}  
& -     
& -     
& -     
& -     
& -     
& -     
& - \\
          
& \textsc{\sffamily cesm-atm} 
& -     
& -     
& -     
& -     
& 47.6 		
& 27.7      
& 17.5 \\
          
& \textsc{\sffamily hurricane} 
& -     
& -     
& -     
& -     
& 83.7 		
& 27.7 		
& 20.8 \\
          
& \textsc{\sffamily nyx}   
& -     
& -     
& -     
& -     
& 71.3		
& 56.3		
& 31.7 \\
          
& \textsc{\sffamily qmcpack} 
& -     
& -     
& -     
& -     
& 72.6		
& 42.5		
& 26.8 \\
\bottomrule

\end{tabular}
\quad
\begin{tabular}{@{} r rr @{}}   
  \grandTableTitle{r}{huffman}{\color{black}\scshape decoding}
& \grandTableTitle{r}{reversed}{\bfseries\scshape(p+q)}
& \grandTableTitle{r}{kernel}{\scshape\hskip -1em decompression}
   \\ 
\toprule
  \TabUnitStyle mb/s
& \TabUnitStyle mb/s
& \TabUnitStyle mb/s
\\
	196.0
&	659.3
& 	151.1
\\
	502.2
&	451.9
&	237.9
\\
	524.5
&	306.8
&	185.0
\\
	670.4
&	300.5
&	201.8
\\
	660.3
&	313.4
&	211.1
\\
\midrule
\scshape canonical
\\[-.4ex]
  \textsc{dec.\hspace{.66em}\TabUnitStyle gb/s}
& \TabUnitStyle gb/s
& \TabUnitStyle gb/s
\\
  35.0	& 16.8	& 11.4	\\
  41.6	& 58.5	& 24.3	\\
  34.2	& 43.9	& 19.2	\\
  52.4	& 29.7	& 19.0	\\
  40.3	& 22.4	& 14.4	\\
\midrule
  - & - & - 	\\ 
  - & - & 113.1 	\\ 
  - & - & 102.2 \\ 
  -	& - & 103.1	\\ 
  -	& - & 115.5	\\ 
\bottomrule

\end{tabular}

}
    \caption{Breakdown comparison of kernel performance among CPU-SZ, \textsc{cuSZ}, and cuZFP. Here ``-'' represents for \textsc{n/a}.}
    \label{tab:breakdown-kernel}
    \vspace{-4mm}
\end{table*}

Next, we evaluate the overall compression and decompression performance of \textsc{cuSZ}, as shown in \TAB~\ref{tab:breakdown-kernel}.
We compare \textsc{cuSZ} with cuZFP in terms of the kernel performance and the overall performance that includes the GPU-to-CPU communication cost. 
Note that the performance of cuZFP is highly related to its user-set fixed bitrate according to the previous study~\cite{jin2020understanding}, whereas the performance of \textsc{cuSZ} is hardly affected by the user-set error bound. Therefore, we choose the acceptable fixed bitrate for cuZFP, which generates data distortion (i.e., PSNR of about 85 dB) similar to that of \textsc{cuSZ}, as shown in \TAB~\ref{tab:rate-distortion}.
Also, note that we exclude cuZFP for \texttt{HACC} in \TAB~\ref{tab:breakdown-kernel}, because cuZFP generates fairly low compression quality on 1D \texttt{HACC}. In particular, even when the bitrate is as high as 16, the PSNR is only about 20 dB, which is not usable.
The throughput in \TAB~\ref{tab:breakdown-kernel} is calculated based on the original data size rather than the size of the data transferred between the GPU and CPU.
\TAB~\ref{tab:breakdown-kernel} shows that cuZFP has a higher kernel throughput but lower GPU-to-CPU throughput than does \textsc{cuSZ}. The reason is that \textsc{cuSZ} provides a much higher compression ratio than does cuZFP with the same data distortion.

We note that the overall throughputs of \textsc{cuSZ} and cuZFP are close to each other 
with respect to the CPU-GPU interconnect (16-lane PCIe 3.0) bandwidth in our evaluation. 
Generally speaking, many applications in GPU-based HPC systems generate the data on GPUs, so the compression needs to be directly performed on the data in the GPU memory, and the compressed data currently must be transferred from GPUs to disks through CPUs. Current state-of-the-art CPU-GPU interconnect technologies such as NVLink~\cite{foley2017ultra} can typically provide a theoretical transfer rate of 50 GB/s over two links, while our \textsc{cuSZ}'s compression kernel can provide comparable throughput of about 40 GB/s. Although cuZFP's compression kernel achieves about 70 GB/s, its overall throughput is limited by the CPU-GPU bandwidth of 50 GB/s. So, the data transfer between CPU and GPU is still the bottleneck for high-throughput compression kernels (e.g., not higher than 50 GB/s).
Moreover, the decompression throughput of \textsc{cuSZ} is lower than its compression throughput and that of cuZFP. This is because only coarse-grained chunking can be applied to decompression, as mentioned in \SEC\ref{sec:decompression}. 
Here we argue that the compression throughput is more important than the decompression throughput, because users  use the CPU-SZ mainly to decompress the data for postanalysis and visualization instead of the GPU after the compressed data is transferred and stored to parallel file systems~\cite{use-case-Franck, jin2020understanding}.

We note that \textsc{cuSZ} on the \texttt{CESM-ATM} dataset exhibits much lower performance than on other datasets. This is due to the fact that each field of the \texttt{CESM-ATM} dataset is fairly small ($\sim$25 MB), such that the codebook construction cost turns out to be relatively high compared with other steps for this dataset. In fact, the codebook construction would not be a bottleneck for a relatively large dataset (such as hundreds of MBs per field), which is more common in practice (e.g., \texttt{HACC}, \texttt{Nyx}, \texttt{QMCPACK}).

We also compare the performance of \textsc{cuSZ} with that of the production version of SZ running on a single CPU core and multiple CPU cores.
The parallelization of OpenMP-SZ is achieved by simply chunking the whole data without any further algorithmic optimization (such as our proposed {\termDQ}). In particular, each thread is assigned with a fixed-size block and runs the original sequential CPU-SZ code. The points on the border are handled similar to \textsc{cuSZ} (as shown in \FIG~\ref{fig:chunk-padding}). 
The main differences between OpenMP-SZ and \textsc{cuSZ} are fourfold: \Circled{1} In the proposed {\termDQ}, each point in \textsc{cuSZ} is assigned to a GPU thread, whereas OpenMP-SZ uses a CPU thread to handle a block of data points. \Circled{2} After {\termPOSTQ}, the data are transformed into integers (units of error bound), and all the following arithmetic operations are performed on these integers. Hence \textsc{cuSZ} does not need to handle the errors that are introduced by floating-point operations (e.g., underflow). \Circled{3} OpenMP-SZ does not fully parallelize Huffman coding, whereas \textsc{cuSZ} provides an efficient parallel implementation of Huffman coding on GPU. \Circled{4} OpenMP-SZ  supports only 3D datasets, so in our comparison we use 3D \texttt{Hurricane Isabel} and \texttt{Nyx} and mark \textsc{n/a} for non-3D datasets in \FIG~\ref{fig:kernel-throughput}.
It illustrate the compression and decompression throughput of \textsc{cuSZ} (considering the CPU-GPU communication overhead) and CPU-SZ. Compared with the serial SZ, the overall compression performance can improved by {\cuszToCpuMin} to {\cuszToCpuMax}.
\textsc{cuSZ} also improves the overall performance by {\cuszToOmpMin} to {\cuszToOmpMax} over SZ running with OpenMP on 32 cores.

\subsubsection{\textbf{Compression Quality}}
\label{subs:eval-quality}

We then present the compression quality of \textsc{cuSZ} compared with another advanced GPU-supported lossy compressor---cuZFP---based on the compression ratios and data distortions on the tested datasets.
We use the \textit{peak signal-to-noise ratio (PSNR)} \footnote{PSNR is calculated as $\text{PSNR}= 20\cdot \log_{10}\! \big[(d_{\max}\! - d_{\min} )/\text{RMSE}\big]$, where $N$ is the number of data points and $d_{\max}$/ $d_{\min}$ is the maximal/minimal value. \textit{Root mean squared error} (RMSE) is obtained by $\text{\sffamily sqrt}\big[\frac{1}{N}\textstyle\sum_{i=1}^{N} \left(d_i - d_i^\bullet\right)^2\!\big]$, where $d_i$ and $d_i^\bullet$ refer to the original and decompressed values, respectively.} to evaluate the quality of the reconstructed data. 
The larger the PSNR, the lower reconstructed distortion, hence the more accurate postanalysis.

We compare \textsc{cuSZ} and cuZFP only on two 3D datasets---\texttt{Hurricane Isabel} and \texttt{Nyx}---because the compression quality of cuZFP on the 1D/2D datasets is much lower than that on the 3D datasets . For a fair comparison, we plot the rate-distortion curves for both \textsc{cuSZ} and cuZFP on all the fields of the two datasets and compare their compression quality in PSNR at the same compression ratio.

\begin{figure}[ht]
    \centering
	\includegraphics[width=.95\linewidth]{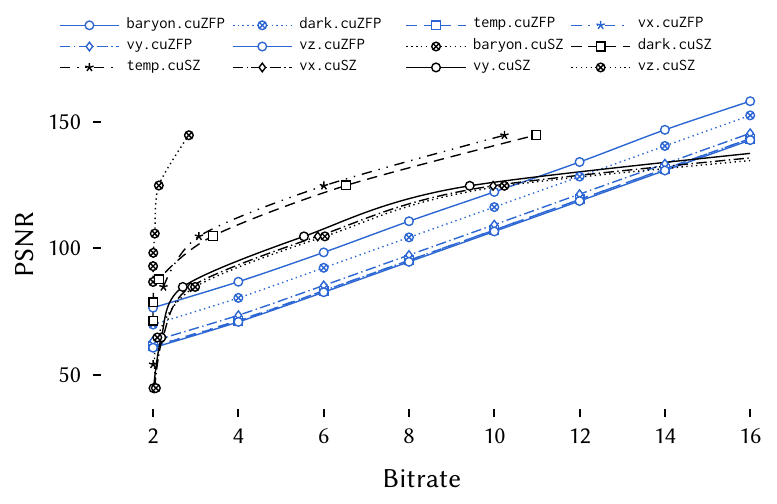}
    \caption{Comparison of rate-distortion between \textsc{cuSZ} (fixed \texttt{valrel}) and \text{cuZFP} (fixed rate) on \texttt{Nyx} dataset.}
    \label{fig:psnr-nyx}
     \vspace{-2mm}
\end{figure}

\FIG~\ref{fig:psnr-nyx} shows the rate-distortion curves of \textsc{cuSZ} and cuZFP on the \texttt{Nyx} dataset. We observe that \textsc{cuSZ} generally has a higher PSNR than does cuZFP with the same compression ratio on the \texttt{Nyx} dataset. In other words, \textsc{cuSZ} provides a much higher compression ratio compared with cuZFP given the same compression quality. The main reason is twofold: \Circled{1} ZFP has better compression quality with the absolute error bound (fix-accuracy) mode
than with the fixed-rate mode (as indicated by the ZFP developer~\cite{fix-accuracy-better-than-fix-rate}); and \Circled{2} the $\ell$-predictor of \textsc{cuSZ} has a higher decorrelation efficiency than does the block transform of cuZFP, especially on the field with a large value range and concentrated distribution, such as \texttt{baryon\_density}.

\begin{figure}[ht]
    \centering
    \includegraphics[width=.95\linewidth]{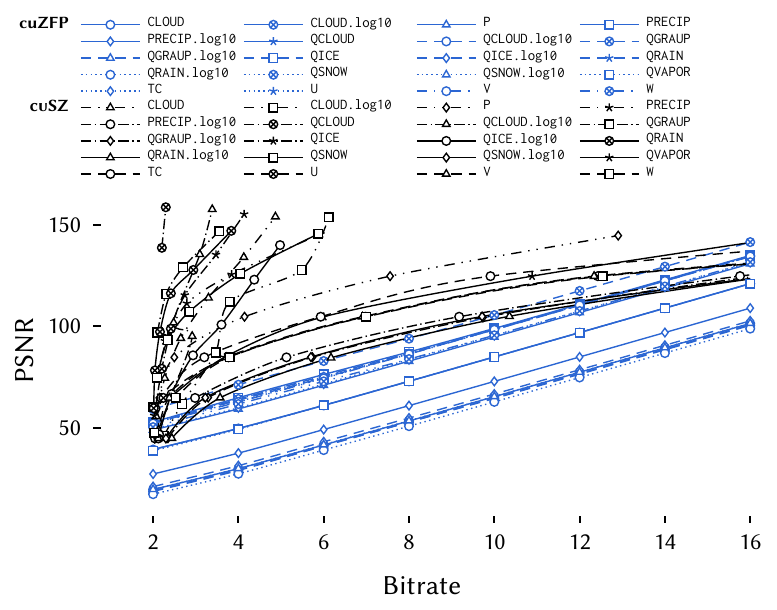}
    \caption{Comparison of rate-distortion between \textsc{cuSZ} (fixed \texttt{valrel}) and \text{cuZFP} (fixed rate) on \texttt{Hurricane Isabel} dataset.}
    \vspace{-2mm}
    \label{fig:psnr-hurricane}
\end{figure}

Similar results for \textsc{cuSZ} and cuZFP are observed on the \texttt{Hurricane Isabel} dataset, as shown in \FIG~\ref{fig:psnr-hurricane}. We note that the rate-distortion curves for \textsc{cuSZ}---namely, \texttt{QCLOUD}, \texttt{QICE}, \texttt{CLOUD}---notably increase when the compression ratio decreases. This is because  there are areas full of zeros, causing the compression ratio to change very slowly when the error bound is smaller than a certain value. In other words, most of the nonzeros are unpredictable, and the zeros are always predictable. 

\begin{figure}[ht]
    \centering
    \includegraphics[width=.95\linewidth]{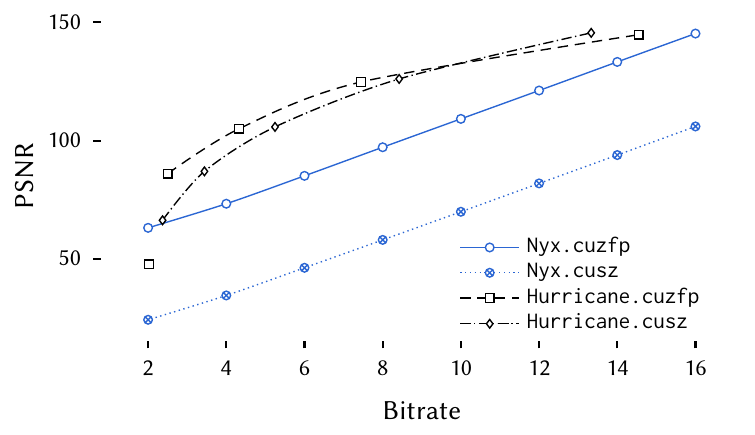}
    \caption{Comparison of overall rate distortion between \textsc{cuSZ} (fixed \texttt{valrel}) and \text{cuZFP} (fixed rate) on \texttt{Hurricane} and \texttt{Nyx} datasets (averaged based on all fields).}
    \vspace{-2mm}
    \label{fig:psnr-overall}
\end{figure}

We also illustrate the overall rate-distortion curves of \textsc{cuSZ} and cuZFP on the \texttt{Hurricane} and \texttt{Nyx} dataset, as shown in \FIG~\ref{fig:psnr-overall}. For example, \textsc{cuSZ} provides a {\cuszToCuzfpBrMin} (2.49 vs. 6) lower bitrate over cuZFP on the \texttt{Nyx} dataset and a {\cuszToCuzfpBrMax} (3.45 vs. 12) lower bitrate over cuZFP on the \texttt{Hurricane Isabel} dataset, with reasonable PSNRs, as shown in \TAB~\ref{tab:rate-distortion}.

\begin{table}[ht]
  \centering
  \scriptsize
  \resizebox{\linewidth}{!}{
  \sffamily
\begin{tabular}{@{}lrrlrr@{}}
\singleLineTableHeadFormat field     & \textsc{\sffamily\bfseries SZ-1.4} & \textsc{\sffamily\bfseries cuSZ} &   \singleLineTableHeadFormat field    & \textsc{\sffamily\bfseries SZ-1.4} & \textsc{\sffamily\bfseries cuSZ} \\
\toprule
    \texttt{CLOUDf48} & 84.99 & 94.18 & \texttt{QSNOWf48} & 84.31 & 93.36 \\
    \texttt{CLOUDf48.log10} & 84.51 & 87.17 & \texttt{QSNOWf48.log10} & 84.87 & 84.93 \\
    \texttt{Pf48} & 84.79 & 84.79 & \texttt{QVAPORf48} & 84.79 & 84.80 \\
    \texttt{PRECIPf48} & 85.35 & 92.86 & \texttt{TCf48} & 84.79 & 84.79 \\
    \texttt{PRECIPf48.log10} & 84.82 & 84.77 & \texttt{Uf48} & 84.79 & 84.79 \\
    \texttt{QCLOUDf48} & 85.03 & 98.91 & \texttt{Vf48} & 84.79 & 84.79 \\
    \texttt{QCLOUDf48.log10} & 85.22 & 95.21 & \texttt{Wf48} & 84.79 & 84.79 \\
\addlinespace[-0.7ex]
\cmidrule(l){4-6}
\addlinespace[-0.8ex]
\texttt{QGRAUPf48} & 88.21 & 97.02 & \texttt{baryon\_density} & 89.71 & 98.25 \\
    \texttt{QGRAUPf48.log10} & 84.90 & 84.82 & \texttt{dark\_matter\_density} & 86.57 & 87.77 \\
    \texttt{QICEf48} & 84.61 & 95.51 & \texttt{temperature} & 84.77 & 84.77 \\
    \texttt{QICEf48.log10} & 85.56 & 85.77 & \texttt{velocity\_x} & 84.77 & 84.77 \\
    \texttt{QRAINf48} & 85.36 & 97.37 & \texttt{velocity\_y} & 84.77 & 84.77 \\
    \texttt{QRAINf48.log10} & 84.93 & 84.56 & \texttt{velocity\_z} & 84.77 & 84.77 \\
\midrule
    \sffamily Hurricane avg. & 85.01 & 86.96 & \sffamily Nyx avg. & 85.58 & 85.98 \\
    \bottomrule
\end{tabular}%
  }
  \caption{Comparison of PSNR between \textsc{cuSZ} and SZ-1.4 on  \texttt{Hurricane} (\textsc{first} 20) and \texttt{Nyx} (\textsc{last} 6) under \texttt{valrel} = $10^{-4}$.}
  \vspace{-6mm}
  \label{tab:sim-to-cpusz}
\end{table}%

\begin{table}[ht]
\centering\scriptsize\sffamily
{\setlength{\tabcolsep}{4pt}  
\renewcommand{\arraystretch}{1.2}

\begin{tabular}{@{}|lllllll|l|@{}}
\multicolumn{8}{l@{}}{\bfseries CLOUDf48} \\
\hline
min & 1\%   & 25\%  & 50\%  & 75\%  & 99\%  & max & range \\
0.00e+0 & 0.00e+0 & 0.00e+0 & 0.00e+0 & 0.00e+0 & 2.53e-4 & 2.05e-3 & 2.05e-3 \\
\hline
\multicolumn{8}{@{}|l| @{}}{\makebox[6.5em][l]{$\phantom{\frac{1}{10}}eb=$2.05e-7}\quad 89.20\% in $[-eb, eb]$, \hspace{2.63em}and 89.20\% in $[\text{min}, \text{min} + eb]$} \\
\multicolumn{8}{@{}|l| @{}}{\makebox[6.5em][l]{$\frac{1}{10}eb=$2.05e-8}\quad 88.50\% in $[-\frac{1}{10}eb, \frac{1}{10}eb]$, and 88.50\% in $[\text{min}, \text{min} + \frac{1}{10}eb]$} 
\\[.4ex]
\hline
\multicolumn{8}{l@{}}{\bfseries QSNOWf48} \\
\hline
min & 1\%   & 25\%  & 50\%  & 75\%  & 99\%  & max & range \\
0.00e+0 & 0.00e+0 & 1.11e-10 & 1.96e-9 & 6.34e-9 & 6.01e-5 & 8.56e-4 & 8.56e-4 \\
\hline
\multicolumn{8}{@{}|l| @{}}{\makebox[6.5em][l]{$\phantom{\frac{1}{10}}eb=$8.56e-8}\quad 88.90\% in $[-eb, eb]$, \hspace{2.63em}and 88.90\% in $[\text{min}, \text{min} + eb]$} \\
\multicolumn{8}{@{}|l| @{}}{\makebox[6.5em][l]{$\frac{1}{10}eb=$8.56e-9}\quad 80.90\% in $[-\frac{1}{10}eb, \frac{1}{10}eb]$, and 80.90\% in $[\text{min}, \text{min} + eb]$} 
\\[.4ex]
\hline
\multicolumn{8}{l@{}}{\bfseries baryon density} \\
\hline
min & 1\%   & 25\%  & 50\%  & 75\%  & 99\%  & max & range \\
5.80e-2 & 1.37e-1 & 3.22e-1 & 5.06e-1 & 8.75e-1 & 7.42e+0 & 1.16e+5 & 1.16e+5 \\
\hline
\multicolumn{8}{@{}|l| @{}}{\makebox[6.5em][l]{$\phantom{\frac{1}{10}}eb=$1.16e+1}\quad 99.50\% in $[-eb, eb]$, \hspace{2.63em}and 99.50\% in $[\text{min}, \text{min} + eb]$} \\
\multicolumn{8}{@{}|l| @{}}{\makebox[6.5em][l]{$\frac{1}{10}eb=$1.16e+0}\quad 83.30\% in $[-\frac{1}{10}eb, \frac{1}{10}eb]$, and 84.40\% in $[\text{min}, \text{min} + \frac{1}{10}eb]$} 
\\[.4ex]
\hline
\end{tabular}%
}

%

\caption{Statistical information (percentile) of example fields having high PSNR under \texttt{valrel} = $10^{-4}$. The range of $eb$ or even $\frac{1}{10}eb$ at \texttt{0} or \text{min} value cover a majority of data in the fields.}
\vspace{-8mm}
\label{tab:nearzero}
\end{table}

The reason is that, according to \SEC\ref{subs:brute-force-par}, \textsc{cuSZ} sets all the values in the padding layer to \texttt{0} and uses these zeros to predict the top-left data points, resulting in better prediction on the tested datasets, especially for the fields with large value ranges and a large majority of values close to zero (such as \texttt{CLOUDf48}, \texttt{QSNOWf48}, and \texttt{baryon\_density} as shown in \TAB~\ref{tab:nearzero}).
However, SZ-1.4's prediction highly depends on the first data point's value, so it may cause low prediction accuracy when the first data point deviates largely  from most of the other points. 
Therefore, \textsc{cuSZ} and SZ-1.4 have similar PSNRs on the datasets represented by the logarithmic scale.

\section{Related Work}
\label{sec:related}

\subsection{GPU-Accelerated Scientific Compression}
\label{sub:gpucx}

Scientific data compression has been studied for many years for reducing storage and I/O overhead. It includes two main categories: lossless compression and lossy compression. Lossless compressors for scientific datasets such as FPC~\cite{fpc} and FPZIP~\cite{lindstrom2006fast} ensure that the decompressed data is unchanged, but they  provide only a limited compression ratio because of the significant randomness of the ending mantissa bit of HPC floating-point data. According to a recent study~\cite{son2014data}, the compression ratio of lossless compressors for scientific datasets is generally up to 2:1, which is much lower than the user-desired ratio for HPC applications.

Error-bounded lossy compression significantly reduces the size of scientific data while maintaining desired data characteristics. Traditional lossy compressors (such as JPEG~\cite{jpeg}) are designed for image and visualization purposes; however, they are difficult to be applied to scientific datasets because of scientists' specific data fidelity requirement. Recently, error-bounded lossy compressors (such as SZ~\cite{sz17} and ZFP~\cite{zfp}) have been developed for scientific datasets. Such compressors provide strict error controls according to user requirements. Both SZ and ZFP, for example, provide an absolute error bound in their CPU version.

Different from SZ's prediction-based compression algorithm, ZFP's algorithm is based on  a block transform. It first splits the whole dataset into many small blocks.
It then compresses the data in each block separately in four main steps: exponent alignment, customized near-orthogonal transform, fixed-point integer conversion, and bit-plane-based embedded coding. A truncation is performed based on the user-set bitrate. Recently, the ZFP team released their CUDA version, called cuZFP~\cite{cuZFP}. cuZFP provides much higher throughputs for compression and decompression compared with the CPU version~\cite{jin2020understanding}. However, the current cuZFP only supports fixed-rate mode, which significantly limit its adoption in practice.

\subsection{Huffman Coding on GPU}
During the Huffman coding process, a specific method is used to determine the bit representation for each symbol, which results in variable length prefix codes.
The set of these prefix codes make up the codebook, with each prefix code  based on the symbols frequency in the data. This codebook is then used to replace each input symbol with its corresponding prefix code.
Previous studies have shown that Huffman coding achieves better performance in parallel on a GPU than in serial on a CPU. 
In general, parallel Huffman coding obtains each codeword from a lookup table (generated by a Huffman tree) and concatenates codewords together with other codewords. 
However, a severe performance issue arises when different threads write codewords with different lengths, which results in warp divergence on GPU \cite{xiang2014warp}. 
The most deviation between methods occurs in concatenating codewords. 

Fuentes-Alventosa et al. \cite{fuentes2014cuvle} proposed a GPU implementation of Huffman coding using CUDA with a given table of variable-length codes, which improves the performance by more than 20$\times$ compared with a serial CPU implementation. 
Rahmani et al. \cite{Rahmani_Topal_Akinlar_2014} proposed a CUDA implementation of Huffman coding based on serially constructing the Huffman codeword tree and parallel generating the byte stream, which can achieve up to 22$\times$ speedups compared with a serial CPU implementation without any constraint on the maximum codeword length or data entropy. 
Lal et al. \cite{Lal_Lucas_Juurlink_2017} proposed a Huffman-coding-based memory compression for GPUs (called E\textsuperscript{2}MC) based on a probability estimation of symbols.
It uses an intermediate buffer to reduce the required memory bandwidth. In order to place the codeword into the correct memory location, E\textsuperscript{2}MC extends the codeword to the size of the buffer length and uses a barrel shifter to write the codeword to the correct location. Once shifted, the codeword is bitwise ORed with the intermediate buffer, and the write location is increased by the codeword length.

\section{Conclusion and Future Work}
\label{sec:conclusion}

In this work, we propose \textsc{cuSZ}, a high-performance GPU-based lossy compressor for NVIDIA GPU architectures that effectively improves the compression throughput for SZ compared with the production version on CPUs.
We propose a dual-quantization scheme to completely remove the strong data dependency in SZ's prediction-quantization step and implement an efficient customized Huffman coding.
We also propose a series of techniques to optimize the performance of \textsc{cuSZ}, including fine-tuning the chunk size, adaptively selecting Huffman code representation, and reusing memory.
Experiments on five real-world HPC simulation datasets show that our proposed \textsc{cuSZ} improves the compression throughput by {\cuszToCpuMin} {to} {\cuszToCpuMax} over the serial CPU version and {\cuszToOmpMin} {to} {\cuszToOmpMax} over the parallel CPU version. 
Compared with another state-of-the-art GPU-supported lossy compressor, \textsc{cuSZ} improves the compression ratio by {\cuszToCuzfpBrMin} to {\cuszToCuzfpBrMax} with reasonable data distortion on the tested datasets. 
We plan to further optimize the performance of decompression, implement other data prediction methods such as linear-regression-based predictor, and evaluate the performance improvements of parallel I/O with \textsc{cuSZ}.

\section*{Acknowledgments}
\footnotesize This research was supported by the Exascale Computing Project (ECP), Project Number: 17-SC-20-SC, a collaborative effort of two DOE organizations---the Office of Science and the National Nuclear Security Administration, responsible for the planning and preparation of a capable exascale ecosystem, including software, applications, hardware, advanced system engineering and early testbed platforms, to support the nation's exascale computing imperative. The material was supported by the U.S. Department of Energy, Office of Science, under contract DE-AC02-06CH11357. This work was also supported by the National Science Foundation under Grants CCF-1619253, OAC-2003709, OAC-1948447/2034169, and OAC-2003624/202042084. We would like to thank The University of Alabama for providing the startup funding for this work.

\renewcommand*{\bibfont}{\small}
\printbibliography[]

\end{document}